\documentclass[fleqn,usenatbib]{mnras}

\usepackage{newtxtext,newtxmath}

\usepackage[T1]{fontenc}
\usepackage{ae,aecompl}

\usepackage{graphicx}	
\usepackage{amsmath}	
\usepackage{amssymb}	
\usepackage{comment}
\usepackage{soul}
\usepackage{xcolor}

\newcommand{\cmfst}{\textsc{21cmfast}}
\newcommand{\cmmc}{\textsc{21cmmc}}
\newcommand\lsim{\mathrel{\rlap{\lower4pt\hbox{\hskip1pt$\sim$}}
        \raise1pt\hbox{$<$}}}
\newcommand\gsim{\mathrel{\rlap{\lower4pt\hbox{\hskip1pt$\sim$}}
        \raise1pt\hbox{$>$}}}

\title[21-cm and luminosity functions]{Inferring the astrophysics of reionization and cosmic dawn from galaxy luminosity functions and the 21-cm signal}

\author[J. Park et al.]{
Jaehong~Park,$^{1}$\thanks{E-mail: jaehong.park@sns.it (JP)}
Andrei Mesinger,$^{1}$\thanks{andrei.mesinger@sns.it (AM)}
Bradley Greig,$^{2,3}$
and Nicolas Gillet$^{1}$
\\
$^{1}$Scuola Normale Superiore, Piazza dei Cavalieri 7, I-56126 Pisa, Italy\\
$^{2}$ARC Centre of Excellence for All-Sky Astrophysics in 3 Dimensions (ASTRO 3D), University of Melbourne, VIC 3010, Australia\\
$^{3}$School of Physics, The University of Melbourne, Parkville, VIC 3010, Australia
}

\date{Accepted XXX. Received YYY; in original form ZZZ}

\pubyear{2018}

\hypersetup{draft}
\begin{document}
\label{firstpage}
\pagerange{\pageref{firstpage}--\pageref{lastpage}}
\maketitle

\begin{abstract}
The properties of the first galaxies, expected to drive the  Cosmic Dawn (CD) and the Epoch of Reionization (EoR), are encoded in the 3D structure of the cosmic 21-cm signal. Parameter inference from upcoming 21-cm observations promises to revolutionize our understanding of these unseen galaxies.  However, prior inference was done using models with several simplifying assumptions.  Here we introduce a flexible, physically-motivated parametrization for high-$z$ galaxy properties, implementing it in the public code {\cmfst}. In particular, we allow their star formation rates and ionizing escape fraction to scale with the masses of their host dark matter halos, and directly compute inhomogeneous, sub-grid recombinations in the intergalactic medium.  Combining current {\it Hubble} observations of the rest-frame UV luminosity function (UV LFs) at high-$z$ with a mock 1000h 21-cm observation using the Hydrogen Epoch of Reionization Arrays (HERA), we constrain the parameters of our model using a Monte Carlo Markov Chain sampler of 3D simulations, {\cmmc}.  We show that the amplitude and scaling of the stellar mass with halo mass is strongly constrained by LF observations, while the remaining galaxy properties are constrained mainly by 21-cm observations.  The two data sets compliment each other quite well, mitigating degeneracies intrinsic to each observation.  All eight of our astrophysical parameters are able to be constrained at the level of $\sim 10\%$ or better.  The updated versions of {\cmfst} and {\cmmc}  used in this work are publicly available.
\end{abstract}




%
%
\section{Introduction}\label{sec:intro}
The birth of the first luminous sources in our Universe heralded the end of the cosmic Dark Ages.  This so-called Cosmic Dawn (CD) culminated in the final phase transition of hydrogen in the intergalactic medium (IGM): the Epoch of Reionization (EoR).  Understanding these cosmic epochs is key to understanding the properties of the first structures of our Universe.  Unfortunately, it is likely that the bulk of the first galaxies are too faint to be observed directly, even with upcoming space-based telescope such as James Webb Space Telescope (JWST; \citealp{Gardner2006}). Luckily, these unseen objects can be studied indirectly through their imprint in the IGM, using the  redshifted 21-cm line.  The 21-cm line from neutral hydrogen can map the ionization and thermal state of the IGM well into the infancy of the CD, making it a revolutionary probe of the early Universe \citep[e.g.][]{Hogan&Rees1979,Scott&Rees1990,Gnedin&Ostriker1997,Madau1997,Shaver1999,Tozzi2000,Gnedin&Shaver2004,FOB2006,Morales&Wyithe2010,Pritchard&Loeb2012}.

For the past decade extensive efforts to detect the 21-cm signal have been made.  These include global (average) signal experiments such as the Experiment to Detect the Global EoR Signature (EDGES\footnote{http://loco.lab.asu.edu/edges\newline https://www.haystack.mit.edu/ast/arrays/Edges}; \citealp{Bowman&Rogers2010}), the Shaped Antenna measurement of the background RAdio Spectrum (SARAS\footnote{http://www.rri.res.in/DISTORTION/saras.html}; \citealp{Patra2013}), Sonda Cosmol{\'o}gica de las Islas para la Detecci{\'o}n de Hidr{\'o}geno Neutro (SCI-HI; \citealp{Voytek2014}), the Large-aperture Experiment to detect the Dark Age (LEDA\footnote{http://www.tauceti.caltech.edu/leda/}; \citealp{Price2018}) and Broadband Instrument for Global HydrOgen ReioNisation Signal (BIGHORNS; \citealp{Sokolowski2015}). Indeed,  \cite{Bowman2018} recently detected a feature in the global signal at $z\approx17$, though concerns remain about its interpretation \citep[e.g.][]{Hills2018}.  Ongoing interferometer experiments  aiming to detect the power spectrum of the signal include the Murchison Wide Field Array (MWA\footnote{https://www.haystack.mit.edu/ast/arrays/mwa/}; \citealp{Bowman2013,Tingay2013}), Low Frequency Array (LOFAR\footnote{http://www.lofar.org}; \citealp{van_Haarlem2013}), Precision Array for Probing Epoch of Reionisation (PAPER\footnote{http://eor.berkeley.edu}; \citealp{Parsons2010}). In the near future, next generation interferometers, such as Hydrogen Epoch of Reionization Array (HERA\footnote{http://reionization.org}; \citealp{DeBoer2017}) and Square Kilometre Array(SKA\footnote{https://astronomers.skatelescope.org}; \citealp{Mellema2013,Koopmans2015}),
will allow us to measure the spatial fluctuations of the 21-cm signal over a wider range of redshift with higher signal to noise. With these instruments we should eventually have 3D maps of the first billion years of our Universe!

This upcoming wealth of 21-cm data will allow us to constrain the bulk properties of the underlying galaxies at a hitherto unseen level of precision. Current EoR observations can provide some insight into the general evolution of reionization (e.g. \citealp{Choudhury&Ferrara2006, Mitra2011,Bouwens2015, Price2016,Greig&Mesinger2017,Gorce2018}). However, the sheer volume of upcoming 21-cm data, and the fact that the signal is sensitive to both the ionization and thermal state of the IGM could usher in a new era of precision astrophysical cosmology using standard Bayesian frameworks.  Bayesian 21-cm parameter inference has been made using on-the-fly sampling of 3D simulations \citep{21CMMC,Greig2017,Greig2018}, as well as interpolating simulations over a parameter grid \citep{Mesinger2014,Mesinger2016,Pober2014,Liu2016,Fialkov2017,Hassan2017}. Recently, neural networks have also been used to predict parameters from 21-cm power spectra \citep{Shimabukuro&Semelin2017}, emulate simulations to bypass on-the-fly Monte Carlo Markov Chain (MCMC) sampling \citep{Kern2017,Schmit&Pritchard2018}, and to directly recover parameters from 21-cm images \citep{Gillet2018}.

However, these early 21-cm inference studies made several simplifying assumptions about the properties of galaxies and IGM structures. For example, most studies assume the stellar mass of galaxies scales linearly with the mass of the host halo (though see, e.g., \citealt{Hassan2017}), and/or that the ionizing escape fraction is a constant. Analytical approaches \citep[e.g.][]{Mason2015,Mutch2016,Sun&Furlanetto2016} based on observations of luminosity functions (LFs) as well as hydrodynamic simulations \citep[e.g.][]{Gnedin&Kaurov2014,Paardekooper2015,Ocvirk2016,Xu2016,Kimm2017,Katz2018}, suggest that both of these quantities have a more complex dependence on the halo properties.
Prior studies also made simplifying assumptions about the role of IGM recombinations and how feedback suppresses star formation in small mass halos.

Motivated by observations of high-redshift galaxy LFs, here we generalize the astrophysical parameterization used in the 21-cm modelling code {\cmfst}\footnote{https://github.com/andreimesinger/21cmFAST}. We allow both the stellar mass and the escape fraction to have a power-law scaling with the mass of the host dark matter halo. Moreover, we directly compute sub-grid, inhomogeneous recombinations, following the approach of \citet{SM14}, removing the often-used yet ad-hoc ionzing photon horizon parameter, $R_{\rm mfp}$. The resulting 8-parameter astrophysical model is both physically-motivated and flexible enough to accommodate a large variaty of galaxy formation scenarios.

We show how current LF observations strongly inform the scaling of star formation rate (SFR) with halo mass; however, they leave most of the remaining galaxy parameters unconstrained even with the addition of reionization observables from the CMB and QSO spectra. Using a Monte Carlo Markov Chain sampler of 3D simulations, {\cmmc}\footnote{https://github.com/BradGreig/21CMMC}, we present parameter forecasts for HERA as an upcoming 21-cm interferometer in combination with current LF observations. We show how the strong synergy between the two observations can result in most of the astrophysical parameters being constrained to the level of $\sim10\%$, or better.

The outline of this paper is as follows. We begin in Section~\ref{sec:model} by describing the astrophysical model including our new empirical parametrization of the galaxy properties. In Section~\ref{sec:LF} we present the UV LFs resulting from our model.  In Section~\ref{sec:21cm} we compute a mock 21-cm observation for a fiducial parameter choice. We briefly summarize our MCMC sampler of 3D simulations, {\cmmc} in Section~\ref{sec:21cmmc}. In Section~\ref{sec:results} we show the resulting constraints on our astrophysical parameters, using the observed UV LFs and the mock 21-cm signal, both individually and combining the data sets.  Finally, we summarize our results in Section~\ref{sec:conclusion}. We assume a standard ${\rm \Lambda}$CDM cosmology based on {\it Planck} 2016 result \citep{PlanckXIII}: ($h$, $\Omega_{\rm m}$, $\Omega_{\rm b}$, $\Omega_{\Lambda}$, $\sigma_{8}$, $n_{\rm s}$)=(0.678, 0.308, 0.0484, 0.692, 0.815, 0.968).  Unless stated otherwise, we quote all quantities in comoving units.

%
%
\section{Astrophysical Model}\label{sec:model}
In this section we introduce a new parametrization for the star formation rate, ionizing escape fraction and their scaling with the mass of the host dark matter halos.  We stress that our simple model does not directly follow individual galaxy evolution, making it only applicable for {\it an ensemble average} of the galaxy population, residing in halos of a given mass.  We note that only $\gsim$ 10 Mpc 21-cm structures (i.e. ionized and heated regions) are large enough to be detected even with SKA; these structures likely form with the combined effort of $\sim$100 -- 1000 sources.  Therefore, the implicit ensemble averaging below is reasonably well justified.

\subsection{Galaxy UV properties}\label{sec:galUV}

We start with the common assumption that the stellar mass of a galaxy, $M_{\ast}$, can be related to the mass of the host halo, $M_{\rm h}$ \citep{Kuhlen2012,Dayal2014,Behroozi2015,Mitra2015,Sun&Furlanetto2016,Mutch2016,Yue2016}:
\begin{equation}\label{eq:Mstar}
M_{\ast}(M_{\rm h}) = f_{\ast} \left( \frac{\Omega_{\rm b}}{\Omega_{\rm m}}\right) M_{\rm h},
\end{equation}
where $f_{\ast}$ is the fraction of galactic gas in stars.  Consistent with observations of the faint galaxy population (e.g. see \citealp{Behroozi2015}), we take $f_{\ast}$ to have a  power-law dependence on the dark matter halo mass\footnote{For the purposes of modeling reionization, we do not care about the massive halos which host AGN bright enough to quench star formation.  These halos are far too rare at high redshifts to contribute to reionization (see e.g. \citealp{Bouwens2015})}:
\begin{equation}\label{eq:F_STAR}
f_{\ast}(M_{\rm h}) = f_{\ast,10} \left( \frac{M_{\rm h}} {10^{10}{\rm M}_{\sun}} \right)^{\alpha_{\ast}},
\end{equation}
where $f_{\rm \ast,10}$ is the fraction of galactic gas in stars normalized to the value in halos of mass $10^{10}\,{\rm M_{\sun}}$ and $\alpha_{\ast}$ is the power-law index.  We impose a physical upper limit of $f_{\ast} \leq 1$.

We assume the SFR can be expressed on average as the total stellar mass divided by a characteristic time-scale\footnote{
      Applying the chain rule, ${\rm d} M_\ast / {\rm d} t = ({\rm d} M_\ast / {\rm d} M_{\rm h}) ({\rm d} M_{\rm h} / {\rm d} t)$, and noting that the Hubble time scales with cosmic time during matter domination, it can be shown that our model implies a halo growth rate of $\dot{M}_{\rm h} \propto M_{\rm h}^{\alpha_\ast} t^{-1}$.  In the future we will consider relaxing this relation, allowing an arbitrary scaling with time, if this is motivated by high-$z$ data.}:
\begin{equation}\label{eq:SFR}
\dot{M_{\ast}}(M_{\rm h},z) =  \frac{M_\ast}{t_\ast H(z)^{-1}},
\end{equation}
where $H(z)^{-1}$ is the Hubble time and the star formation time-scale, $t_{\ast}$, is a free parameter which we allow to vary between zero and unity.\footnote{Note that this is the only redshift dependence of our model.  All other parameters are assumed to be functions of only the halo mass.  We note that having the star formation scale with the Hubble time is analogous to having it scale with the dynamical time of DM halos, $t_{\rm dyn}\sim 1/\sqrt{{\rm G}\rho} \sim 1/\sqrt{180{\rm G}\bar{\rho}(z)}$, where $\sim$180 is the mean overdensity of a halo in the spherical collapse model and $\bar{\rho}(z)$ is the background density. Since at high-$z$ the Universe is matter dominated, we have $\bar{\rho} = \rho_{\rm crit} = 3H(z)^2 / 8\pi{\rm G}$, thus $t_{\rm dyn} \propto H(z)^{-1}$.
}

Similarly, we allow the escape fraction, $f_{\rm esc}$, to scale with halo mass according to: 
\begin{equation}\label{eq:F_ESC}
f_{\rm esc}(M_{\rm h}) = f_{{\rm esc, 10}}\left( \frac{M_{\rm h}}{10^{10}{\rm M}_{\sun}}\right)^{\alpha_{\rm esc}},
\end{equation}
where $f_{\rm esc, 10}$ is the normalization of the ionizing UV escape fraction and $\alpha_{\ast}$ is the power-law scaling of $f_{\rm esc}$ with halo mass. $\alpha_{\rm esc}$ is likely to be negative, as SNe can more easily evacuate low column density channels from shallower DM potentials; the escape of ionizing photons is thought to be determined by the covering fraction of these low column density channels \citep[e.g.][]{Paardekooper2015,Xu2016,Katz2018}. As for the stellar fraction, we impose a physical upper limit of $f_{\rm esc} \leq 1$.

Star formation in small galaxies is expected to be quenched due to SNe feedback, photo-heating feedback, or inefficient gas accretion \citep[e.g][]{Shapiro1994,Giroux1994,Hui1997,Barkana&Loeb2001,Springel&Hernquist2003,Okamoto2008,Mesinger2008,Sobacchi2013a,Sobacchi2013b}. We account for this suppression with a redshift-independent duty cycle.
\begin{equation}\label{eq:DC}
f_{\rm duty}(M_{\rm h}) = \exp\left( - \frac{M_{\rm turn}}{M_{\rm h}}\right).
\end{equation}
The duty cycle in our model can be thought of as the fraction of halos of a given mass, which host stars/galaxies.  It is unity for larger halo masses, but as one approaches smaller halos, not all of them will be hosting galaxies.  A fraction $[1 - \exp(- M_{\rm turn}/M_{\rm h})]$ of halos are unable to host a star-forming galaxy, likely because they have not accreted gas efficiently or because a negligible amount of prior star formation sterilized the halo through feedback.  The remaining $\exp(- M_{\rm turn}/M_{\rm h})$ of halos manage to host stars with an efficiency $f_\ast$.  This is something expected from theory, as one approaches cooling/feedback thresholds, and is also seen in hydrodynamic simulations (e.g. fig. 3 in \citealt{O'Shea2015}; left panel of fig. 22 in \citealt{Xu2016}; Gillet et al. in prep).

\subsection{Galaxy X-ray properties}\label{sec:galXray}

By heating the IGM prior to the bulk of reionization, X-rays are expected to play a dominant role during the CD \citep[e.g.][]{Pritchard&Furlanetto2007,McQuinn&O'Leary2012,Mesinger2013}. For each simulation cell at a given spatial position $\mathbf{x}$, and redshift $z$, {\cmfst} computes the angle-averaged specific X-ray intensity (in units of ${\rm erg\,s^{-1}\,keV^{-1}\,cm^{-2}\,sr^{-1}}$), by integrating the comoving X-ray emissivity, $\epsilon_{\rm X}(\mathbf{x},E,z)$, back along the light-cone:
\begin{equation}\label{eq:specific_intensity}
J(\mathbf{x},E,z) = \frac{(1+z)^3}{4\upi}\int_{z}^{\infty}{\rm d}z'\frac{{\rm c}\,{\rm d}t}{{\rm d}z'}\epsilon_{\rm X}{\rm e}^{-\tau},
\end{equation}
where $e^{-\tau}$ accounts for attenuation by hydrogen and helium in the IGM (see equation 16 of \citealt{21cmfast}). The comoving specific emissivity is calculated in the emitted frame, $E_{\rm e}=E(1+z')/(1+z)$, and is given by
\begin{equation}\label{eq:specific_emissivity}
\epsilon_{\rm X}(\mathbf{x},E_{\rm e},z') = \frac{L_{\rm X}}{{\rm SFR}} \left[ (1+\bar{\delta}_{\rm nl})\int_{0}^{\infty}{\rm d}M_{\rm h}\frac{{\rm d}n}{{\rm d}M_{\rm h}}f_{\rm duty}\dot{M_{\ast}} \right],
\end{equation}
where $\bar{\delta}_{\rm nl}$ is the mean, non-linear overdensity of the shell around $({\bf x},z)$, and the term inside square brackets corresponds to the star formation rate density along the light-cone. The conditional halo mass function (HMF), $\frac{{\rm d}n}{{\rm d}M_{\rm h}}(M_{\rm h}, z | R, \delta_{\rm R})$, is obtained by normalizing the conditional Press-Schechter HMF \citep{Lacey&Cole1993,Somerville&Kolatt1999} so as to have the mean of the Sheth-Tormen HMF \citep{S-T1999,S-T2002}, as discussed in \citet{21cmfast} (see also \citealt{Barkana&Loeb2004,Barkana&Loeb2008}). Here the galaxy duty cycle, $f_{\rm duty}(M_h)$ (see equation~\ref{eq:DC}) accounts for inefficient star formation inside small mass halos, and the SFR, $\dot{M}_\ast(M_h, z)$, depends on halo mass and redshift as specified in equation~(\ref{eq:SFR}). 

The term, $L_{\rm X}/{\rm SFR}$, is the specific X-ray luminosity per unit star formation escaping the host galaxies in units of ${\rm erg\,s^{-1}\,keV^{-1}\,M_{\sun}^{-1}\,yr}$. We assume the specific luminosity follows a power-law in photon energy, i.e. $L_{\rm X} \propto E^{-\alpha_{\rm X}}$, and adopt $\alpha_{\rm X} = 1$, consistent with models and observations of local high-mass X-ray binaries (HMXBs) over the relevant energy range \citep[e.g.][]{Fragos2013}. We normalize the specific X-ray luminosity using the integrated soft-band ($< 2{\rm keV}$) luminosity per SFR (in units of ${\rm erg\,s^{-1}\,M_{\sun}^{-1}\,yr}$),
\begin{equation}\label{eq:soft_X-ray}
L_{\rm X<2\,keV}/{\rm SFR} = \int_{E_0}^{2\,{\rm keV}}{\rm d}E_{\rm e}\, L_{\rm X}/{\rm SFR},
\end{equation}
where $E_0$ is the X-ray energy threshold below which photons are absorbed inside the host galaxies. This X-ray energy threshold depends on the density of the interstellar medium (ISM) and metallicity \citep{Das2017}.  The upper limit of the integral is motivated by the fact that the mean free path of $\sim 2$ keV photons is roughly the Hubble length at these redshifts; thus  harder photons do not contribute to IGM heating during the CD (e.g. \citealt{McQuinn2012}).

The soft-band luminosity from eq. (\ref{eq:soft_X-ray}) is kept as a free parameter, while we fix the spectral index, $\alpha_X$, to unity.  Keeping the spectral index constant is motivated by the fact that the 21-cm power spectra (PS) are very insensitive to the spectral index when using this parametrization (e.g. see figure~1 in \citealp{Greig2017}).

\subsection{Inhomogeneous IGM recombinations}\label{sec:rec}

Recombinations can impact the progress and topology of reionization via the interplay of ionizing sources and dense IGM structures (so-called Lyman limit systems). If reionization is ``photon-starved" as suggested by emissivity estimates from the Lyman alpha forest, recombinations would ``stall" the growth of large $\ion{H}{II}$ regions \citep[e.g.][]{Miralda-Escude2000,Ciardi2006,McQuinn2007,Finlator2012,Kaurov2014,SM14}. In semi-numerical simulations, this effect is usually crudely accounted for with a maximum horizon for ionizing photons (commonly denoted $R_{\rm mfp}$) which is usually taken to be redshift independent and homogeneous \citep[e.g.][]{21cmfast,Alvarez&Abel2012,Greig&Mesinger2017}.

Here we directly compute the local, sub-grid recombinations, according to \cite{SM14} (see also \citealp{Hutter2018}). Specifically, each simulation cell at a spatial location ${\bf x}$ and redshift, $z$, keeps track of its hydrogen recombination rate according to:
\begin{equation}\label{eq:Dn_recDt}
\frac{{\rm d}n_{\rm rec}}{{\rm d}t}(\mathbf{x},z) = \bar{n}_{\rm H} \alpha_{\rm B} \Delta_{\rm cell}^{-1} \int_0^{180} \left[ 1-x_{\ion{H}{I}}\right]^2 P_{\rm V} \Delta^2 {\rm d}\Delta\,,
\end{equation}
where $\Delta_{\rm cell} \equiv 1+\delta_{\rm nl}$ is the overdensity on the size of the simulation cell, $\Delta \equiv n / \bar{n}$ is the sub-grid overdensity, $P_{\rm V}(\Delta,\Delta_{\rm cell}, z)$ is the volume-averaged PDF of $\Delta$ (with the functional form specified by \citealp{Miralda-Escude2000} and adjusted for the cell's overdensity according to \citealp{SM14}), $\alpha_{\rm B}$ is the case-B recombination coefficient evaluated at a temperature of 10$^4$ K, and $x_{\ion{H}{I}}(\Delta, \Gamma, z)$ is the neutral fraction at the overdensity $\Delta$ with the attenuation of the local, inhomogeneous ionizing background $\Gamma$ accounted for using the analytic expression from \cite{Rahmati2013}.  The upper limit of the integral is motivated by the mean density of halos in the spherical collapse model, since by definition recombinations inside galaxies are accounted for in the source terms. However in practice this limit is unimportant at the redshifts of interest as gas already starts to self-shield at much lower densities for realistic models of $\Gamma$; thus large densities do not contribute to IGM recombinations.

The reionization field in {\cmfst} is then computed by comparing the cumulative number of ionizing photons in a given region of scale $R$ to the corresponding number of baryons {\it plus} the average, cumulative number of recombinations inside that region:
\begin{equation}\label{eq:n_rec}
\bar{n}_{\rm rec}(\mathbf{x},z, R) = \left< \int_{z_{\rm ion}}^{z} \frac{{\rm d}n_{\rm rec}}{{\rm d}t}\frac{{\rm d}t}{{\rm d}z} {\rm d}z\right>_R
\end{equation}
where $z_{\rm ion}({\bf x})$ is the redshift at which a given cell was first ionized. For more details, see section~3 in \cite{SM14}.

\subsection{Summary of the free parameters in our model}\label{sec:free_params}

Our new model has eight free parameters. Here we summarize these parameters, also listing them in Table~\ref{Table:free_parameters}, together with the fiducial values and allowed ranges for the MCMC.  We stress that the fiducial values are only used when we generate a mock 21-cm observation (see \S 6.2); for the MCMC using UV luminosity functions (\S 6.1) we take currently-available observations.

\begin{enumerate}
\item {$f_{\ast,10}$}: the normalization of the fraction of galactic gas in stars at high-$z$, $f_{\ast}$, evaluated for halos of mass $10^{10} {\rm M_{\sun}}$.Our fiducial value used to generate a mock 21-cm signal is $f_{\ast,10} = 0.05$ and we allow the parameter to vary in range of ${\rm log_{10}}(f_{\ast,10})=[-3,0]$.

\item $\alpha_{\ast}$: the power-law scaling of $f_{\ast}$ with halo mass.  When making a mock 21-cm observation, we take a fiducial value of $\alpha_\ast = 0.5$ \citep[e.g.][]{Behroozi2015,Ocvirk2016,Mirocha2017} and we allow the parameter to vary in range of $\alpha_{\ast}=[-0.5,1]$ in our MCMCs.

\item $f_{\rm esc,10}$: the normalization of the ionizing UV escape fraction of high-$z$ galaxies, $f_{\rm esc}$, evaluated for halos of mass $10^{10}{\rm M_{\sun}}$. When making a mock 21-cm observation, we take a fiducial value of $f_{\rm esc,10} = 0.1$ and for our MCMCs we allow the parameter to vary in range of ${\rm log_{10}}(f_{\rm {esc,10}})=[-3,0]$.

\item $\alpha_{\rm esc}$: the power law scaling of $f_{{\rm esc}}$ with halo mass. We take a fiducial value of $\alpha_{\rm esc} = -0.5$.  As mentioned earlier, we expect $\alpha_{\rm esc}$ to be negative as SNe can more easily evacuate low column density channels from shallower potential wells \cite[e.g.][]{Razoumov2010,Yajima2011,Ferrara&Loeb2013,Paardekooper2015,Xu2016,Kimm2017,Katz2018}.
We allow the parameter to vary in range of $\alpha_{\rm esc}=[-1,0.5]$.

\item $t_{\ast}$: the star formation time-scale taken as a fraction of the Hubble time, $H^{-1}(z)$. We take a fiducial value of $t_\ast = 0.5$ and we allow the parameter to vary in range of $t_{\ast}=(0,1]$.

\item $M_{\rm turn}$: the turnover halo mass below which the abundance of active star forming galaxies is exponentially suppressed according to a duty cycle of $\exp( - M_{\rm turn}/M_{\rm h})$. When making a mock 21-cm observation, we take a fiducial value of $M_{\rm turn} = 5\times 10^{8} {\rm M_{\sun}}$ and in the MCMCs we allow the parameter to vary in range of ${\rm log_{10}}(M_{\rm turn})=[8,10]$.  Here the lower limit is motivated by the atomic cooling threshold, while the upper limit is motivated by the faint end of current UV LFs (see Fig.~\ref{fig:21cm_signal}).

\item $E_0$: the minimum X-ray photon energy capable of escaping the galaxy; softer photons are absorbed by the ISM of high-z galaxies.  Motivated by the hydrodynamic simulations used in \citet{Das2017}, we take a fiducial value of $E_0= 0.5 {\rm keV}$ and we allow the parameter to vary in range of $E_0=[0.1,1.5]$.  Analogously, this range corresponds to ${\rm log_{10}}(N_{\ion{H}{I}}/{\rm cm^2})=[19.3,23.0]$.\footnote{The conversion to column densities is computed assuming a unity optical depth for a metal-free column of neutral ISM.} 

\item $L_{\rm X<2\,keV}/{\rm SFR}$: the normalization of the soft-band X-ray luminosity per unit star formation, computed over the band $E_0$ -- 2 keV.  When making a mock 21-cm observations, we assume the X-ray binary composite SED of \cite{Fragos2013}, with the ISM attenuation from \citet{Das2017}, resulting in a fiducial value of $L_{\rm X<2\,keV}/{\rm SFR} = 10^{40.5} {\rm erg\,s^{-1}M_{\sun}^{-1}yr}$.  In our MCMCs, we allow the parameter to vary in the range ${\rm log_{10}}(L_{\rm X<2\,keV}/{\rm SFR})=[38,42]$.
\end{enumerate}

\begin{table}
\begin{center}
\caption{
The astrophysical parameters of our model, together with the fiducial values used for the mock 21-cm signal, and the allowed range for the MCMC studies. We also note the choice of prior in the final column: flat in either linear or log space within the quoted range.  See \S.~\ref{sec:free_params} for additional details.
                  }
                  \vspace{-0.5cm}
\begin{tabular} {ccccc}
\\
\hline\\[-3.0mm]
 Parameter                                                  & Fiducial    & Units        &  Allowed  & Flat prior \\
  & value & & range & \\[0.5mm] \hline\\[-2.5mm]
$f_{\ast,10}$                       & $0.05$      & --                      &  0.001 -- 1 &   log    \\[0.5mm]
$\alpha_{\ast}$                                              & $0.5$    & --                      &  -0.5 -- 1   &   linear  \\[0.5mm]
$f_{\rm esc,10}$                    &  0.1    & --                     &  0.001 -- 1   &   log    \\[0.5mm]
$\alpha_{\rm esc}$                                      & $-0.5$     & --                      &  -1 -- 0.5  &   linear  \\[0.5mm]
$M_{\rm turn}$                      & $5 \times 10^8$ & $M_\odot$ &  $ 10^8$ -- $10^{10}$     &   log \\[0.5mm] 
 $t_{\ast}$                                                       & $0.5$   & --                       &  0 -- 1     &   linear   \\[0.5mm] 
 $\frac{L_{{\rm X}<2{\rm keV}}}{\rm SFR}$  & $10^{40.5}$ & ${\rm erg\,s^{-1}\,M_{\sun}^{-1}\,yr}$  & $10^{38}$ -- $10^{42}$ &   log\\[0.5mm] 
 $E_{0}$                                                          & 0.5              & {\rm keV} & 0.1 -- 1.5 &   linear\\[0.5mm]  
  \hline
  
\end{tabular}
\label{Table:free_parameters}
\end{center}
\end{table}

\section{Corresponding UV luminosity functions}\label{sec:LF}
 
 We can write the non-ionizing UV luminosity functions (LF) from our model as:
\begin{equation}\label{eq:LF}
\phi(M_{\rm UV}) = \left[ f_{\rm duty} \frac{{\rm d}n}{{\rm d}M_{\rm h}} \right] \left|\frac{{\rm d}M_{\rm h}}{{\rm d}M_{\rm UV}}\right| ~ ,
\end{equation}
where as previously noted, the term in brackets is the number density of active, star-forming galaxies.  The final term on the RHS encodes the conversion of halo mass to UV magnitude.  We evaluate this assuming that the SFR is proportional to the rest-frame UV luminosity of a galaxy:
\begin{equation}\label{eq:L_UV}
\dot{M_{\ast}}(M_{\rm h}, z) = \mathcal{K}_{\rm UV} \times L_{\rm UV}.
\end{equation}
We assume the conversion factor, $\mathcal{K}_{\rm UV}$,  is constant and adopt $\mathcal{K}_{\rm UV} = 1.15 \times 10^{-28} {\rm M_{\rm \sun}\,yr^{-1}/ergs\,s^{-1}\,Hz^{-1}}$ following \citet{Sun&Furlanetto2016}\footnote{This conversion depends on the IMF, as well as the dust content of the galaxy.  In the above, we ignore dust extinction for the faint galaxies and high redshifts of interest (e.g. \citealt{Bouwens2012,Capak2015}), and assume a Salpeter IMF.  However, we note that these uncertainties can roughly be subsumed in our $f_\ast$ parameter.} \citep[see also e.g.][]{Madau1998,Kennicutt1998,Bouwens2012}. Finally we relate the UV luminosity to magnitude using the usual AB magnitude relation \citep{Oke&Gunn1983},
\begin{equation}\label{eq:M_UV}
{\rm log_{10}}\left( \frac{L_{\rm UV}}{{\rm erg\,s^{-1}\,Hz^{-1}}} \right) = 0.4 \times (51.63 - M_{\rm UV}).
\end{equation}

\begin{figure*}
\begin{center}
\includegraphics[width=18cm]{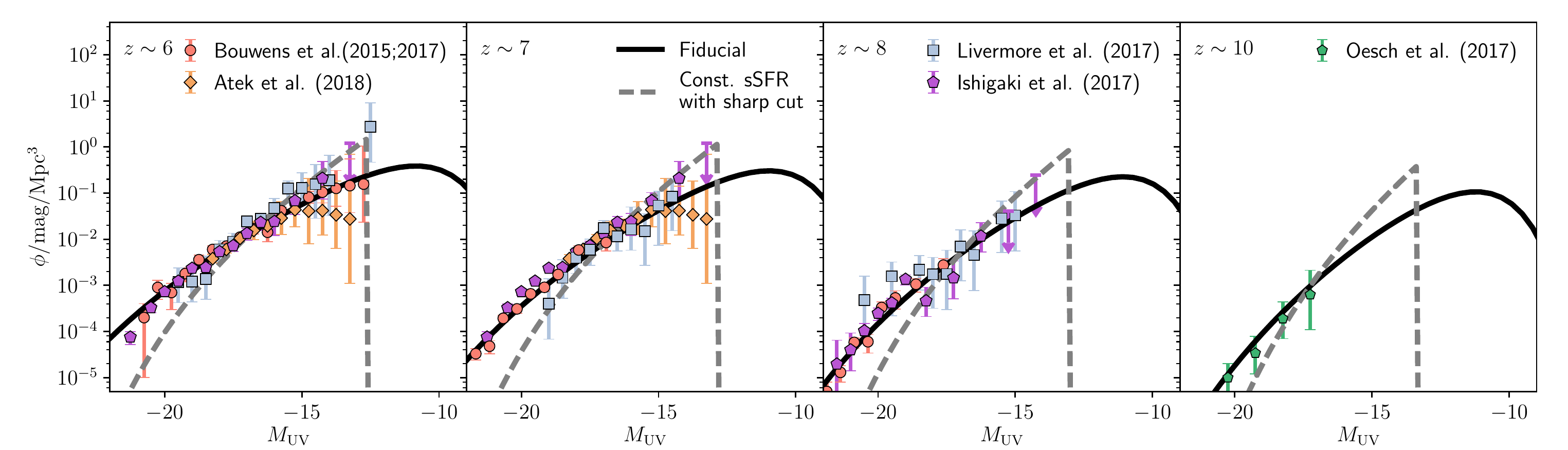}
\end{center}
\vspace{-3mm}
\caption{
Luminosity functions corresponding to our fiducial parameters used to make the mock 21-cm observation below ({\it black solid lines}), together with observed LFs ({\it colored points}).  Note that LFs of \protect\cite{Ishigaki2018} and \protect\cite{Atek2018} are duplicated at $z\sim6$ and $7$, because their selection criteria ($i$-dropout) select intermediate redshift $z\sim 6-7$ galaxies. 
The dashed curves illustrate common simplifications made in previous 21-cm studies: (i) fixing $\alpha_\ast = 0$, thus setting a constant stellar mass to halo mass for all galaxies; and (ii) having a sharp suppression of faint galaxies (i.e. with $f_{\rm duty}$ transitioning from 1 to 0 at $M_{\rm turn}$).
}
\label{fig:LF}
\end{figure*}

For illustration purposes, in Fig.\,\ref{fig:LF} we show the rest-frame UV LFs corresponding to the fiducial model parameters we use to make mock 21-cm observations (Table 1), as well current observations of UV LFs.  The observations show scatter between various groups, in particular at the faint end which is dominated by lensing uncertainties  \citep{Bouwens2015_LF4-10,Bouwens2016_LF6,Livermore2017_LF,Atek2018,Ishigaki2018}.   Our fiducial parameter choices are consistent with current observations.  More important than the fiducial parameter choices is the flexibility of our model to reproduce the main features of the LFs, {\it within a physical framework based on the HMFs}.  It is important to note that empirical models of LF  which are not directly rooted in the assumption that galaxies sit in halos (such as the Schechter function) require ad hoc tuning to capture the redshift evolution inherent to the HMF.  Indeed \cite{Oesch2018} show that simple estimates based on the HMF can more accurately predict very high-redshift star formation rates, compared with empirical ones.

\begin{figure}
\begin{center}
\includegraphics[width=9.5cm]{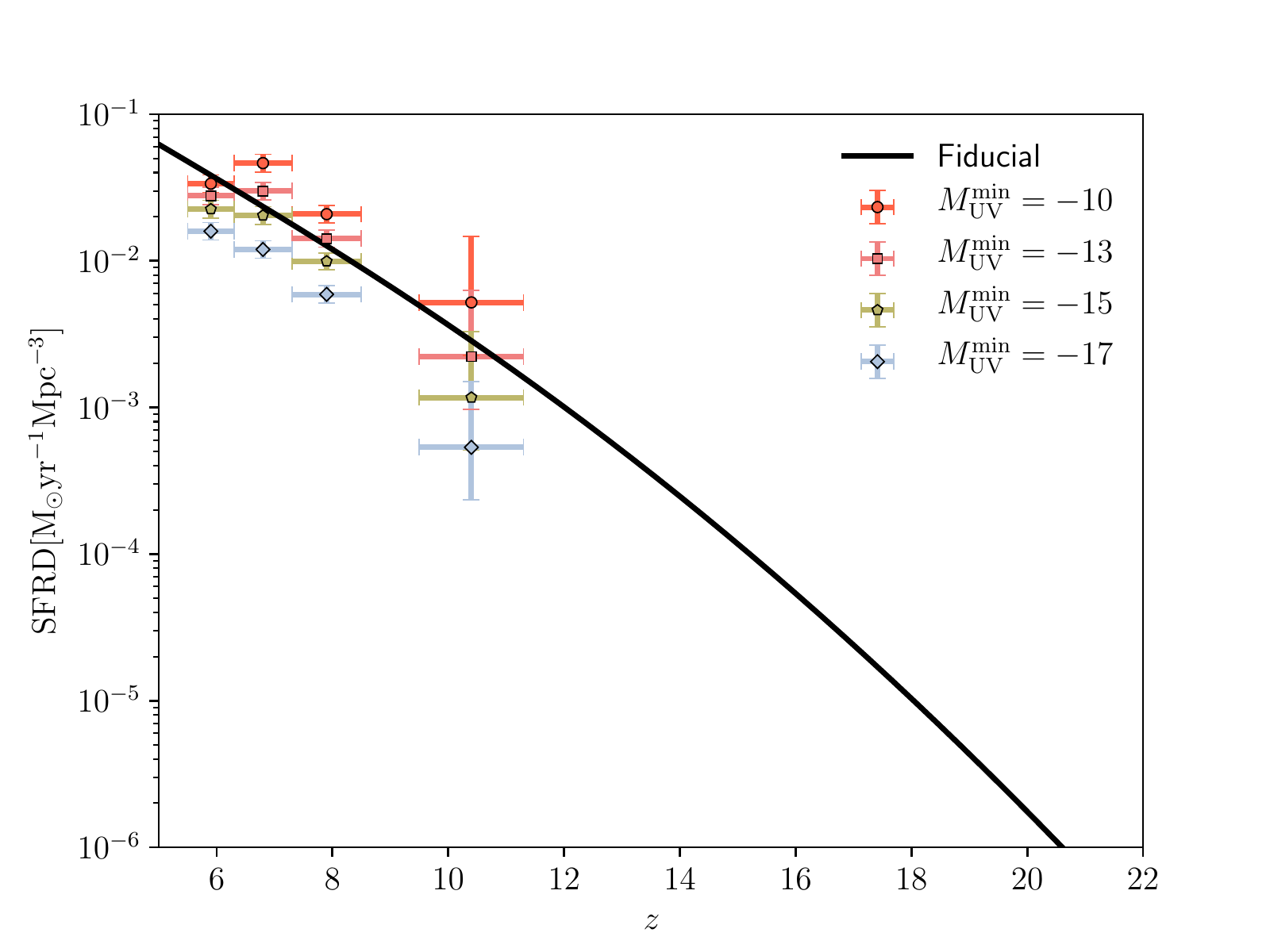}
\end{center}
\vspace{-3mm}
\caption{
Star formation rate density for our fiducial model ({\it solid line}).  Circles, squares, pentagons and diamonds represent estimates from \protect\citep{Bouwens2015} obtained by extrapolating LFs down to minimum magnitudes of $M_{\rm UV}^{\rm min}=-10$, $-13$, $-15$ and $-17$, respectively.
}
\label{fig:SFRD}
\end{figure}

Additionally, in Figure~\ref{fig:SFRD} we show the star formation rate density (SFRD) for our fiducial model parameters (see equation~\ref{eq:specific_emissivity}).  For comparison, we also show the estimates of \cite{Bouwens2015}, obtained by extrapolating their observed LFs down to minimum magnitudes of $M_{\rm UV}^{\rm min}=-17$, $-15$, $-13$ and $-10$, truncating them sharply beyond those values. The SFRD corresponding to our fiducial choice of $M_{\rm turn} = 5 \times 10^8 M_\odot$ is roughly comparable to the sharp cut-off assumption of \cite{Bouwens2015} for $M_{\rm UV}^{\rm min}=-10$ -- -13.

\section{Corresponding 21-cm signal}\label{sec:21cm}

\subsection{Computing the signal}
The 21-cm signal is commonly expressed as the offset of the 21-cm brightness temperature, $\delta T_{\rm b}(\nu)$, relative to the temperature of the cosmic microwave background (CMB), $T_{\rm CMB}$ \citep[e.g.][]{Furlanetto2006}:
\begin{equation}\label{eq:T_brtness}
\begin{split}
\delta T_{\rm b} &\approx 27 x_{\ion{H}{I}} (1 + \delta_{\rm nl}) \left(\frac{H}{{\rm d}v_{\rm r}/{\rm d}r + H}\right)\left(1 - \frac{T_{\rm CMB}}{T_{\rm S}}\right)\\
                 &\quad \times \left( \frac{1+z}{10} \frac{0.15}{\Omega{\rm m}h^2} \right)^{1/2} \left(\frac{\Omega_{\rm b}h^2}{0.023} \right),
\end{split}
\end{equation}
where $ x_{\ion{H}{I}}$ is the neutral fraction, $T_{\rm S}$ is the gas spin temperature, $\delta_{\rm nl} \equiv \rho/{\bar \rho}-1$ is the gas overdensity, $H(z)$ is the Hubble parameter, ${\rm d}v_{\rm r}/{\rm d}r$ is the gradient of the line-of-sight component of the velocity and all quantities are evaluated at redshift $z=\nu_0/\nu - 1$, where $\nu_0$ is the 21-cm frequency. 

To compute the various fields in the above equation, we use the semi-numerical simulation {\cmfst}\citep{Mesinger2007,21cmfast}. {\cmfst} computes the evolved density and velocity fields from an initial high resolution Gaussian realization, using second order LPT \citep[e.g.][]{Scoccimarro1998}. Then, {\cmfst} estimates the ionization field from the evolved IGM density field by comparing the cumulative number of ionizing photons to the number of neutral atoms plus cumulative recombinations within spheres of decreasing radii (e.g. \citealt{Furlanetto2004}). Specifically, a voxel at coordinates $(\mathbf{x},z)$ is flagged as fully ionized if it satisfies 
\begin{equation}\label{eq:ionized}
n_{\rm ion}(\mathbf{x},z| R, \delta_{\rm R}) \geq (1 + \bar{n}_{\rm rec})(1 - \bar{x}_{\rm e}),
\end{equation}
with $\bar{n}_{\rm rec}$ given by eq. (\ref{eq:n_rec}) and the cumulative number of IGM ionizing photons per baryon produced inside a spherical region of scale $R$ and corresponding overdensity $\delta_{\rm R}$ given by:
\begin{equation}\label{eq:nion}
n_{\rm ion} = \bar{\rho}^{-1}_{\rm b} \int_0^\infty {\rm d}M_{\rm h} \frac{{\rm d}n(M_{\rm h}, z | R, \delta_{\rm R})}{{\rm d}M_{\rm h}} f_{\rm duty} M_{\ast} f_{\rm esc} N_{\rm \gamma/b},
\end{equation}
where $\bar{\rho}_{\rm b}$ is the mean baryon density in the region, $N_{\gamma/{\rm b}}$ is the number of ionizing photons per stellar baryon which we set at 5000, motivated by a Salpeter IMF (in principle this parameter is largely degenerate with $f_\ast$). Note that in our new formulation, the commonly used ``ionizing efficiency" parameter, $\zeta = f_\ast f_{\rm esc} N_{\rm \gamma/b}$, is broken-up into its constituent parts, with $f_\ast(M_h)$ and $f_{\rm esc}(M_h)$ now both being functions of halo mass (as discussed in \S~\ref{sec:galUV}). The final term on the RHS of eq. (\ref{eq:ionized}) is a small correction factor for pre-ionization by X-rays, discussed below; in practice, this term is negligible for realistic models (see \citealt{Mesinger2013}).

The above procedure is used to compute the inhomogeneous topology of reionization, consisting of (almost) fully ionized and neutral regions.  However, due to their long mean free paths, X-ray and soft UV photons are able to penetrate even the neutral cosmic patches distant from galaxies.  These radiation fields help determine the spin temperature.  To calculate $T_S$, {\cmfst} follows the evolution of the ionized fraction inside the neutral IGM, $x_{\rm e}$, the kinetic temperature, $T_{\rm K}$, and the incident Lyman $\alpha$ background. The ionization fraction and the kinetic temperature in each voxel are solved following
\begin{equation}\label{eq:dxe_dz}
\frac{{\rm d}x_{\rm e}(\mathbf{x},z')}{{\rm d}z'} = \frac{{\rm d}t}{{\rm d}z'}\left[ \Gamma_{\rm ion, X} - \alpha_{\rm A}C x_{\rm e} n_{\rm b} f_{\rm H}  \right],
\end{equation}
\begin{equation}\label{eq:dTk_dz}
\begin{split}
\frac{{\rm d}T_{\rm K}(\mathbf{x},z')}{{\rm d}z'} &= \frac{2}{3k_{\rm B}(1+x_{\rm e})} \frac{{\rm d}t}{{\rm d}z'}\sum_{\rm p}Q_{\rm p}\\ 
     &\quad + \frac{2T_{\rm K}}{3n_{\rm b}}\frac{{\rm d}n_{\rm b}}{{\rm d}z'} - \frac{T_{\rm K}}{1+x_{\rm e}}\frac{{\rm d}x_{\rm e}}{{\rm d}z'},
\end{split}
\end{equation}
where $n_{\rm b}$ is the total (H + He) baryonic number density at $(\mathbf{x},z')$, $\epsilon_{\rm p}$ is the heating rate per baryon for process ${\rm p}$ in ${\rm erg\,s^{-1}}$, $\alpha_{\rm A}$ is the case-A recombination coefficient, $C$ is the clumping factor on the scale of the simulation cell, $k_{\rm B}$ is the Boltzmann constant, $f_{\rm H}$ is the hydrogen number fraction, $\Gamma_{\rm ion, X}$ is the ionizing background from X-rays, and $Q_p$ is the heating rate per baryon associated with process ``p"; we include Compton heating and X-ray heating.

The heating and ionization rates per baryon inside the mostly neutral IGM are calculated with \citep[see also, e.g.][]{Baek&Ferrara2013,Madau&Fragos2017,Eide2018} 
\begin{equation}\label{eq:Xray_heating_rate}
{Q_{\rm X}}(\mathbf{x},z) = \int_{{\rm Max}[\nu_0,\nu_{\rm \tau=1}]}^{\infty} {\rm d}\nu \frac{4\pi J}{h\nu} \sum_{i}(h\nu-E_{i}^{\rm th})f_{\rm heat}f_{i}x_{i}\sigma_{i}
\end{equation}
and 
\begin{equation}\label{eq:Gamma_ion}
\begin{split}
\Gamma_{\rm ion, X}(\mathbf{x},z) &= \int_{{\rm Max}[\nu_0,\nu_{\rm \tau=1}]}^{\infty} {\rm d}\nu \frac{4\pi J}{h\nu} \sum_{i}f_{i}x_{i}\sigma_{i}F_{i}\\ 
F_i &= (h\nu-E_{i}^{\rm th}) \left( \frac{f_{\rm ion,\ion{H}{I}}}{E_{\ion{H}{I}}^{\rm th}} + \frac{f_{\rm ion,\ion{He}{I}}}{E_{\ion{He}{I}}^{\rm th}} + \frac{f_{\rm ion,\ion{He}{II}}}{E_{\ion{He}{II}}^{\rm th}} \right) + 1,
\end{split}
\end{equation}
where $i = \ion{H}{I},\,\ion{He}{I},\,\ion{He}{II}$ denotes the atomic species, $f_i$ is the corresponding number fraction, $x_i$ is the ionization fraction of the cell's species, $\sigma_i$ is the ionization cross-section, $E_{i}^{\rm th}$ is the ionization threshold energy of species $i$, and $f_{\rm heat}$ and $f_{{\rm ion},j}$ are the fraction of the primary electron's energy going into heating and secondary ionizations of species $j$, respectively. The angle-averaged specific X-ray intensity $J({\mathbf x},E,z)$ is computed from equations (\ref{eq:specific_intensity}) and (\ref{eq:specific_emissivity}).

With the gas kinetic temperature calculated according to the above equations, the spin temperature can be approximated as a weighted average of the CMB and gas temperatures.  Specifically, we have
\begin{equation}\label{eq:Ts}
T_{\rm S}^{-1} = \frac{T_{\rm CMB}^{-1}+x_{\alpha} T_{\alpha}^{-1}+x_{\rm c} T_{\rm K}^{-1}}{1+x_{\rm \alpha}+x_{\rm c}}.
\end{equation}
Here $T_{\alpha}$ is the color temperature which is closely rated to the gas temperature through multiple Lyman $\alpha$ scatterings \citep{Field1959}. For $T_{\rm S}$ not to be equal to the CMB temperature (and hence for us to obtain a signal), either the collisional coupling coefficient, $x_{\rm c}$, or the Wouthuysen-Field (\citealt{Wouthuysen1952,Field1958}; WF) coupling coefficient, $x_{\alpha}$, need to be non-negligible. The former is only efficient in the IGM at $z\gsim 30$ while the later is set by the Lyman $\alpha$ background.  {\cmfst} computes the Lyman series background from both X-ray excitation of HI and from direct stellar emission of photons in the Lyman bands, using the composite stellar spectra of \cite{Barkana&Loeb2005}. It scales with the star formation rates implied by our model, in a manner analogous to equation (\ref{eq:Gamma_ion}).
For more details on the calculations, interested readers are encouraged to consult \cite{Mesinger2007,21cmfast}.

\subsection{Mock 21-cm observation}
\label{sec:mock}

Using the fiducial parameters listed in Table~\ref{Table:free_parameters}, we generate a mock realization of the 21-cm signal.  Our simulation box is 500 Mpc on a side, computed on a 256$^3$ grid, downsampled from 1024$^3$ initial conditions. 
When performing the MCMC, we create 3D simulations on-the-fly, whose dimensions are 250 Mpc on a side on a 128$^3$ grid.  As in previous works, we use a different random seed (and corresponding density realization) for the mock observation than we do for the MCMC inference below. Power spectra are generated from light-cones, using the approach from \citet{Greig2018}. More details, including the power spectra of our simulations can be found in Appendix \ref{app:Mock}.

\begin{figure*}
\begin{center}
\includegraphics[width=18cm]{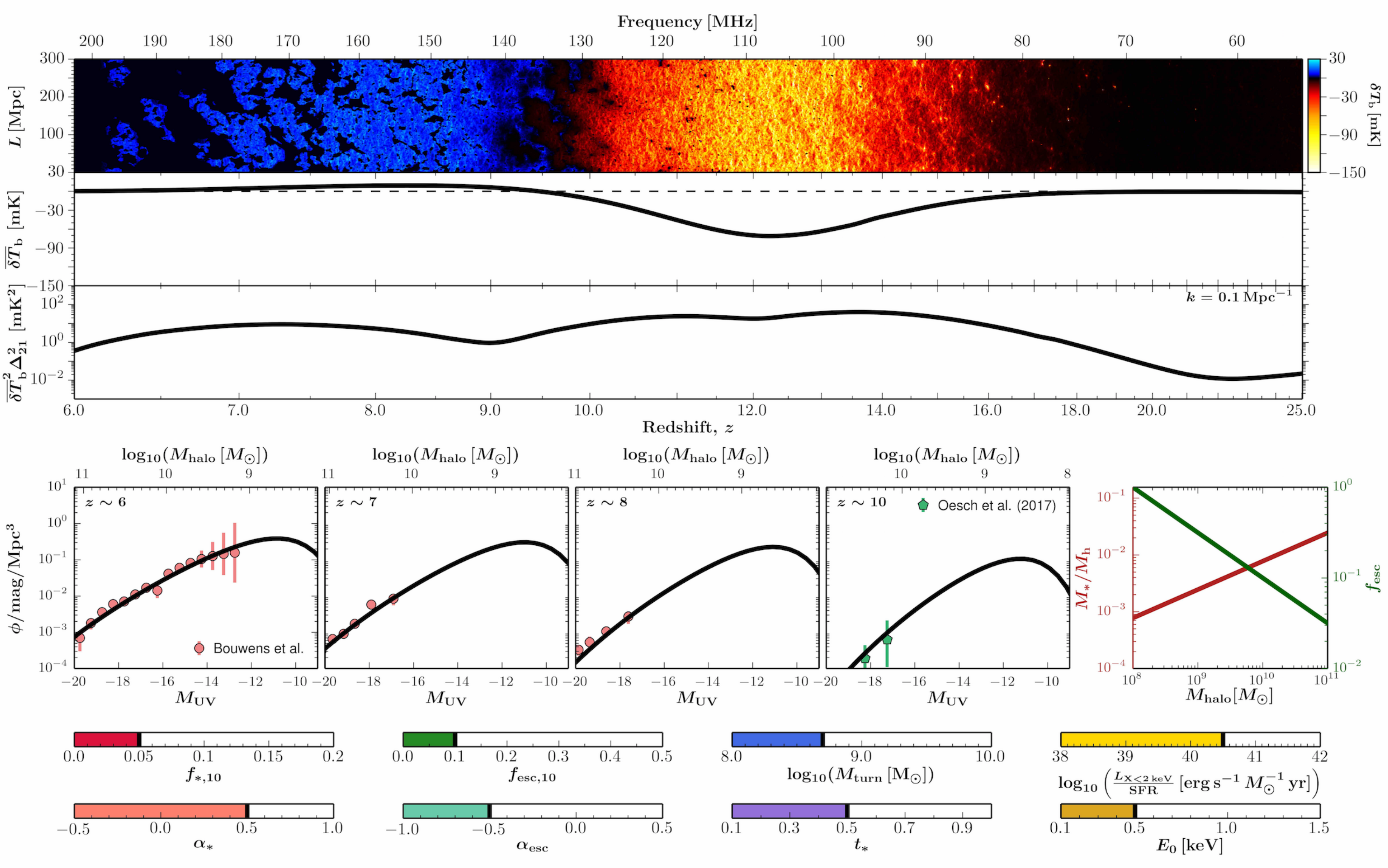}
\end{center}
\vspace{-3mm}
\caption{
The 21-cm signal together with the UV LFs corresponding to our fiducial model parameters.  The top three panels show a $\sim 1$ Mpc slice through the 3D light-cone of 21-cm signal, the average brightness temperature offset and the PS at $k = 0.1\,{\rm Mpc^{-1}}$, respectively. The left four panels in the middle show corresponding LFs with observations from \protect\cite{Bouwens2016_LF6} for $z\sim6$, \protect\cite{Bouwens2015_LF4-10} for $z\sim7-8$ and  \protect\cite{Oesch2018} for $z\sim10$, respectively. The rightmost panel in the middle shows the stellar mass per halo mass ({\it left axis}) and the escape fraction ({\it right axis}) as functions of halo mass. Toggles on the bottom represent the fiducial parameter values. For movies showing how these observables change with changes in the astrophysical parameters, see \url{http://homepage.sns.it/mesinger/Videos/parameter_variation.mp4}
}
\label{fig:21cm_signal}
\end{figure*}

Figure~\ref{fig:21cm_signal} shows a 2D light-cone slice of the 21-cm signal which corresponds to our fiducial model parameters.  A 2D slice through the light cone is shown in the top panel, the average brightness temperature is shown in the second panel, and the PS at $k = 0.1\,{\rm Mpc^{-1}}$ is shown in the third panel.  The corresponding LFs and scalings of the stellar mass per halo mass and the escape fraction are shown in the bottom panels.

For these parameter choices, we obtain an end to reionizaiton which is consistent with current observations \citep[e.g.][]{McGreer2015};  however, the EoH and epoch of WF coupling are delayed compared with our previous works \citep[e.g.][]{Mesinger2016,Greig2018}. Our new fiducial model has the star-formation efficiency, $f_\ast$, increase with the halo mass, rather than remain constant as we had done previously.  The ionizing escape fraction, $f_{\rm esc}$, has the opposite scaling with halo mass for our fiducial choices; as reionization depends on the product of these quantities, the EoR timing is unchanged.  However, the EoH is governed by X-rays, which are unaffected by $f_{\rm esc}$\footnote{Note that in our model, the minimum X-ray energy escaping the ISM is set by the average HI column density of early galaxies, which we take to be {\it independent} from the UV escape fraction, $f_{\rm esc}$.  This is motivated by the emerging physical picture in which the ionizing escape fraction is determined by the covering fraction of low-column density sightlines in early galaxies resulting from feedback events.  This low value tail of the column density distribution (determining $f_{\rm esc}$) is not sensitive to the mean of the distribution (determining $E_0$)\citep[e.g.][]{Verhamme2015,Xu2016,Das2017,Pallottini2017}.}.
As the importance of small mass galaxies increases with redshift, the new scaling of $f_\ast(M_h)$ results in a delayed EoH compared to previous $f_\ast=$ const. We note that  \cite{Mirocha2017} reached a similar conclusion for the global 21-cm signal, using a  similar parametrization to the one we use here. 
Consistent with their follow-up work in \cite{Mirocha&Furlanetto2018}, this implies that if the claim of a detection of the EoH at $z\sim17$ made recently by EDGES \citep{Bowman2018} is proven genuine, star-formation would either need to extend to very small halos below the atomic cooling threshold, and/or star-formation would need to be more efficient in small-mass halos below current LF limits, implying a break in the power-law scaling of $f_\ast(M_{\rm h})$.  Currently we are extending our model to capture an additional and separate population of sources residing in smaller, molecularly-cooled halos, in an effort to quantify these claims (Qin et al., in prep.).

The other notable difference between Figure~\ref{fig:21cm_signal} and previous results is that the EoH and WF coupling epochs are less separated in time.  This is due to two factors: (i) the decreasing star formation efficiency with halo mass, discussed above, resulting in a delayed and subsequently more rapid CD (an effect similar to having the dominant population of star forming galaxies sitting in more massive halos whose fractional abundance evolves more rapidly); and (ii) our fiducial value of $L_{\rm X, <2 keV}/{\rm SFR}$ is larger than in most previous studies.  Our new value is motivated by the theoretical HMXB models of \cite{Fragos2013}, whereas previously we used the empirical scalings of \cite{Mineo2012} obtained from local star forming galaxies.  Due to its dependence on metallicity \citep{Basu-Zych2013,Lehmer2016,Brorby2016}, the X-ray luminosity to SFR for the first galaxies is expected to be roughly an order of magnitude larger than for local ones.

The fact that the epoch of WF coupling and EoH are more coincident in time is evidenced by the smaller separation between the corresponding peaks of the large-scale power, driven by spatial fluctuations in the Ly$\alpha$ coupling coefficient and gas temperature, respectively \citep[e.g.][]{Pritchard&Furlanetto2007}. Moreover, the global absorption signal has a reduced minimum, as the heating commences before all of the IGM has its spin temperature coupled to the gas kinetic temperature.  Similarly, the peak in the power spectrum associated with the EoH is reduced, as the cross-terms from the coupling coefficient and gas temperature have a negative contribution to the power amplitude (see the discussion in \citealp{Pritchard&Furlanetto2007} and \citealp{Mesinger2016}).

%
%
\section{Sampling astrophysical parameter space with 21CMMC}\label{sec:21cmmc}

In this section we provide a summary of {\cmmc} \citep{21CMMC} used to constrain the astrophysical parameters described in section~\ref{sec:free_params}. For further details, interested readers are referred to \cite{21CMMC,Greig2017,Greig2018}.

{\cmmc} is an MCMC sampler of 3D reionzation simulations. To explore the astrophysical parameter space of cosmic dawn and reionization, {\cmmc} adopts a massively parallel MCMC sampler \textsc{cosmohammer} \citep{Akeret2013} that uses the \textsc{emcee python} module \citep{Foreman-Mackey2013} based on the affine invariant ensemble sampler \citep{Goodman2010}. At each proposed MCMC step, {\cmmc} calculates an independent 3D light-cone realization of the 21-cm signal, using an optimized version of {\cmfst}. Then, it calculates a likelihood by comparing PS of the sampled 21-cm signal against the mock observation (see Appendix~\ref{app:Mock}), defined as
\begin{equation}\label{eq:PS}
\delta{\bar T}_{\rm b}^2\Delta_{21}^{2}(k,z) \equiv \frac{k^3}{2\pi^2V}\delta{\bar T}_{\rm b}^2(z) \left< |\delta_{21}({\mathbf k},z)|^2 \right>_{k},
\end{equation}
where $\delta_{21}({\mathbf x},z) \equiv \delta{\bar T}_{\rm b}({\mathbf x},z)/\delta{\bar T}_{\rm b}(z)-1$. Note that we limit the $k$ space range from $0.1$ to $1.0$, corresponding roughly to limits on the foreground noise and the shot noise, respectively

As in previous works,  we adopt a modeling uncertainty, accounting for inaccuracies in our semi-numerical models.  We take a constant uncertainty of 20 per cent on the sampled 21-cm PS, motivated by comparisons to RT simulations \citep{Zahn2011,Ghara2015,Hutter2018}. We note that with further comparisons, these modeling uncertainties can be better characterized and accounted for.
Moreover, we include Poisson uncertainties on the sampled 21-cm PS, roughly consistent with cosmic variance for these scales \citep{Mondal2015}. These two  uncertainties are added in quadrature with the total noise PS in equation~\ref{eq:total_noise}. 

We account for redshift space distortions along the line of sight using the relation
\begin{equation}\label{eq:RSD}
{\mathbf s} = {\mathbf x} + \frac{(1+z)}{H(z)} v_{\parallel}({\mathbf x}),
\end{equation}
where $\mathbf s$ and $\mathbf x$ are the redshift and real space signal, respectively. For details of this implementation, see  \cite{Greig2018} (see also \citealp{Mao2012,Jensen2013}).

\subsection{Telescope noise}\label{sec:noise}

We calculate noise on the mock 21-cm observation using the python module \textsc{21cmsense} \citep{Pober2013,Pober2014}. First, we generate the thermal noise PS at each $uv$ cell according to \citep[e.g.][]{Morales2005,McQuinn2006,Pober2014}:
\begin{equation}\label{eq:thermal_noise}
\Delta_{\rm N}^2(k) \approx X^2Y \frac{k^3}{2\pi^2}\frac{\Omega'}{2t}T_{\rm sys}^2,
\end{equation}
where $XY^2$ is a scalar factor converting observed bandwidths and solid angles to comoving distance, $\Omega'$ is a beam-dependent factor derived in \cite{Parsons2014}, $t$ is the integration time within a particular $k$-mode, $T_{\rm sys}$ is the system temperature. Then, the total noise power at a given Fourier mode $k$, with an assumption of Gaussian errors for the cosmic-variance term, is expressed as
\begin{equation}\label{eq:total_noise}
\delta\Delta_{\rm T+S}^{2}(k) = \left(\sum_{i} \frac{1}{[\Delta_{{\rm N},i}^2(k) + \Delta_{21}^2(k)]^2}\right)^{-\frac{1}{2}},
\end{equation}
where $\Delta_{21}^{2}(k)$ is the 21-cm PS from the mock observation.

For our fiducial instrument, we take the HERA design described in \cite{Beardsley2015}: a core consisting of 331 dishes. We assume a 1000 h observation, spread across 180 nights at 6 hours per night, and an  observing bandwidth coverage of $50 - 250\,{\rm MHz}$.  We note that previous studies using a reduced parameter set have shown comparable constraints with SKA when using the PS statistic (e.g. \citealt{Greig2017}).  However, these claims might need to be re-evaluated for our expanded parametrization.  We postpone this to future work, as this paper is mainly a proof-of-concept for the benefit of combining observables.

%
%
\section{Results: Recovery of Astrophysical parameters}\label{sec:results}

We now quantify how current and upcoming observations are able to constrain our model parameters. We use two main observations: (i) current high-$z$ UV LFs; and (ii) 21-cm power spectra from a mock 21-cm observation described in \S~\ref{sec:mock}. We first quantify the utility of each in turn, before combining them. We also include current EoR constraints, but as we show below, these do not improve parameter constraints beyond what is available with (i) and (ii).  Our results are summarized in Table.~\ref{Table:recovered_parameters}, where we list recovered median values for our model parameters together with $68$ per cent confidence regions for each data set used in the MCMC.

\begin{table*}
\begin{center}
\caption{
         Summary of the median recovered values and $1\,\sigma$ errors for the eight-parameter astrophysical model, obtained from our MCMC procedure for each combination of data sets listed below. The LF observations are from \citet{Bouwens2016_LF6, Bouwens2015_LF4-10, Oesch2018}, the $\tau_e$ constraints are from \protect\cite{Planck2016}, the dark fraction constraints are from \protect\cite{McGreer2015}, while the 21-cm data corresponds to power spectra extracted from a mock 1000h observation with HERA331.}
\begin{tabular} {ccccccccc}
\\
\hline\\[-3.0mm]
               &  &  &  &  Parameters  &  &  &       \\[1mm]
               & ${\rm log_{10}}(f_{\ast,10})$ & $\alpha_{\ast}$ & ${\rm log_{10}}(f_{\rm esc,10})$ & $\alpha_{\rm esc}$ & ${\rm log_{10}}(M_{\rm turn})$ & $t_{\ast}$ & ${\rm log_{10}}\left(\frac{L_{{\rm X}<2{\rm keV}}}{\rm SFR}\right)$ & $E_0$   \\[1mm]
               &  &  &  &  & $[{\rm M_{\sun}}]$  &   & $[{\rm erg\,s^{-1}\,M_{\sun}^{-1}\,yr}]$ &  $[{\rm keV}]$   \\[1.5mm] \hline \\[-2.5mm]
  Fiducial values  & $-1.30$ & $0.50$ & $-1.00$ & $-0.50$ & $8.7$  & $0.5$ & $40.50$ &  $0.50$   \\[1.5mm] \hline\\[-2.5mm]              
  LF only      & $-1.25^{+0.20}_{-0.39}$ &  $0.50^{+0.07}_{-0.06}$ &  .  &  .   &   $8.68^{+0.40}_{-0.41}$   &  $0.51^{+0.30}_{-0.30}$   &    .  &  . \\[1.5mm]
LF + $\tau_{\rm e}$ + the dark fraction & $-1.21^{+0.18}_{-0.30}$ &  $0.50^{+0.07}_{-0.07}$ &  $-0.91^{+0.42}_{-0.35}$ &  $-0.13^{+0.44}_{-0.53}$   &   $8.65^{+0.44}_{-0.41}$   &  $0.55^{+0.28}_{-0.27}$   &    .  &  . \\[1.5mm]
21-cm only     & $-1.29^{+0.18}_{-0.21}$ &  $0.38^{+0.23}_{-0.31}$ &  $-0.99^{+0.24}_{-0.21}$  &  $-0.42^{+0.26}_{-0.27}$   &   $8.80^{+0.27}_{-0.26}$   &  $0.46^{+0.17}_{-0.14}$   &    $40.46^{+0.07}_{-0.07}$  &   $0.50^{+0.04}_{-0.04}$\\[1.5mm]
21-cm $+$ LF   &  $-1.20^{+0.14}_{-0.14}$ &  $0.47^{+0.06}_{-0.06}$  &  $-1.10^{+0.16}_{-0.18}$  &  $-0.48^{+0.14}_{-0.18}$  &   $8.76^{+0.19}_{-0.23}$   &  $0.56^{+0.21}_{-0.16}$   &   $40.49^{+0.05}_{-0.06}$   &   $0.50^{+0.03}_{-0.03}$\\[1.5mm]
  \hline
  
\end{tabular}
\label{Table:recovered_parameters}
\end{center}
\end{table*}

\subsection{Using only galaxy luminosity functions}\label{sec:Constraint-LF}

\begin{figure*}
\begin{center}
\includegraphics[width=17.5cm]{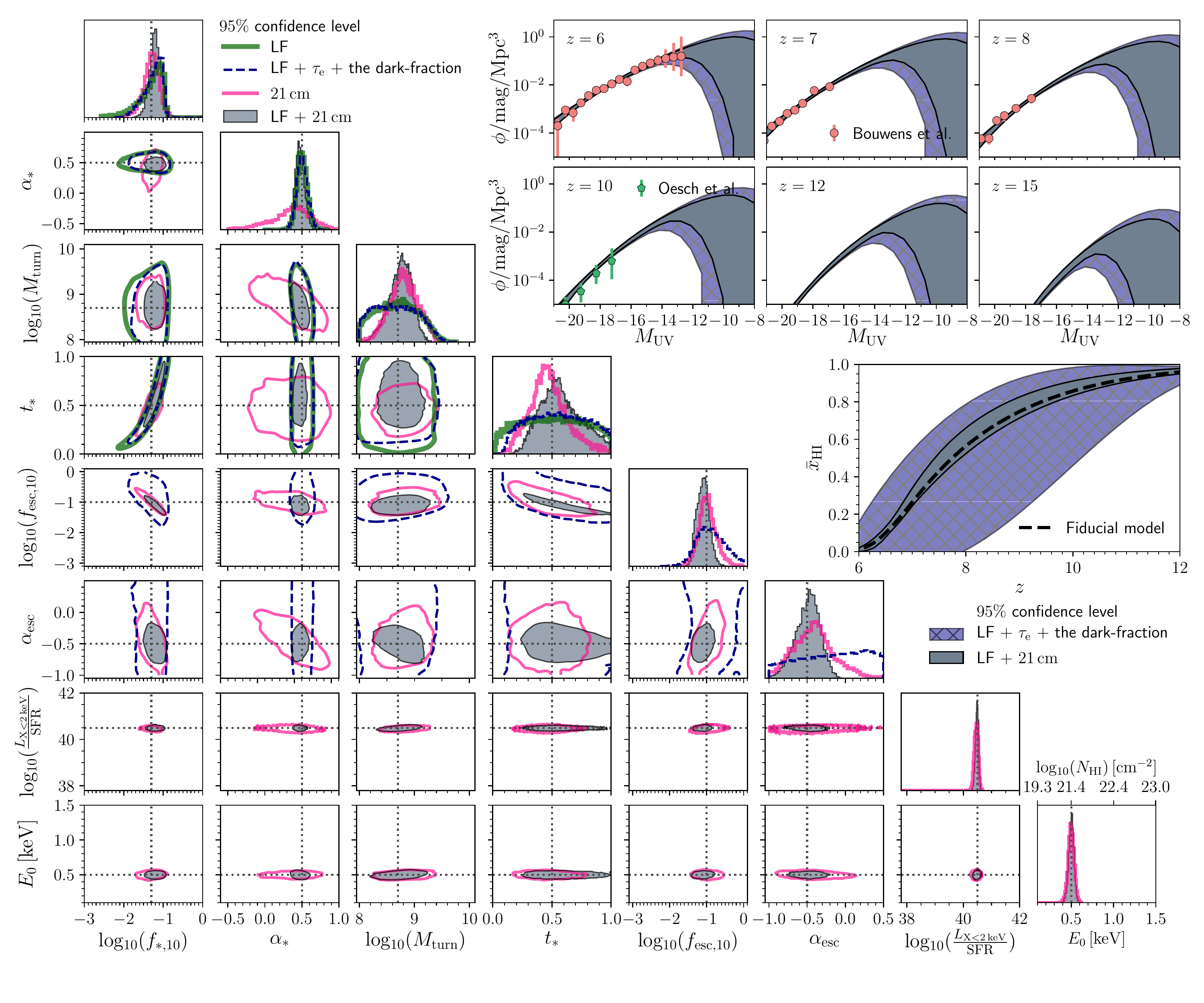}
\end{center}
\vspace{-3mm}
\caption{
  Corner plot showing parameter constrains for the various data sets used (see legend): 2D marginalized joint posterior distributions are shown in the bottom left corner with 1D marginalized PDFs along the diagonal. Green solid lines, blue dashed lines, pink solid lines, and shaded regions represent $95$ per cent confidence levels for constraints using data sets of LF only, LF + $\tau_{\rm e}$  + the dark fraction, the mock 21-cm PS, and both LF + the mock 21-cm PS, respectively. Top-right panels: Recovered $95$ per cent confidence levels of the LFs corresponding to the posterior of our model. Shaded regions with hatch (blue) and shaded regions (grey) represent constraints using LF + $\tau_{\rm e}$  + the dark fraction and using 21-cm with the UV LFs, respectively. Middle-right: Corresponding constraints on the  global evolution of the IGM neutral fraction, $x_{\ion{H}{I}}(z)$. The dashed line represents the fiducial model used to make the mock 21-cm observation. Shaded regions with hatch (blue) and shaded regions (grey) represent the recovered $95$ per cent confidence levels for constraints using LF + $\tau_{\rm e}$  + the dark fraction and using both 21-cm with the UV LFs, respectively. 
}
\label{fig:contour}
\end{figure*}

The LFs from our model depend on four free parameters: $f_{\ast,10}$, $\alpha_{\ast}$, $M_{\rm turn}$ and $t_{\ast}$.  We begin by quantifying how current observations can constrain these parameters.  To compute the likelihood in our MCMC, we use the  $z\sim6$ LFs from \cite{Bouwens2016_LF6}, $z\sim7-8$  LFs from \cite{Bouwens2015_LF4-10} and $z\sim10$ LFs from \cite{Oesch2018}.  Using these data points and corresponding error bars, we compute the $\chi^2$ likelihood at each of the four redshifts, and multiply them together when calculating the posterior. We only use data points at $M_{\rm UV} > -20$, as our model does not account for AGN feedback or dust extinction in bright galaxies since these are far too rare to be relevant for reionization and cosmic dawn.

We stress that we are {\it not} trying to rigorously quantify constraints available with current UV LF observations.  In order to do this properly, one should account for systematic uncertainties, combining the various estimates from different groups, some of which are shown in Fig.~\ref{fig:LF}.  We postpone such an investigation for future work.  By using observations from a single group, we are somewhat overestimating the current constraining power of UV LFs, illustrating their future potential when/if systematics can be better understood and we can have ``concordance" LFs.

The constraints on our four model parameters that determine LFs are denoted with the solid green curves in the triangle plot of figure \ref{fig:contour}, and are summarized in the first row of table \ref{Table:recovered_parameters}.  Given the allowed range of parameter space, the most robust constraints we obtain are on $\alpha_{\ast} = 0.50^{+0.07}_{-0.06} (1 \sigma)$.  This parameter most strongly affects the {\it slope} of the LFs, which are very well determined by current observations (see also very similar conclusions reached by \citealt{Mirocha2017}, who use a similar galaxy formation model).

By contrast, current LFs only set an upper limit for $M_{\rm turn}$, ruling out $M_{\rm turn} \lesssim 9.5$ due to the presence of faint galaxies.  The marginalized 1D PDF below this value is relatively flat, due to the large uncertainties at the faint end, and also to the lack of an identifiable turn over in the observational data sets we use.

Our remaining two galaxy model parameters are only constrained as a ratio, as is evidenced by the strong degeneracy of the green curve in the $f_{\ast, 10}$ -- $t_\ast$ panel of the triangle plot.  This is understandable as the LF in our model is determined only by the SFR, which scales as $f_{\ast} / t_\ast$.

\subsubsection{Do current constraints on the reionization history improve parameter inference and allow us to constrain the escape fraction?}

The 1500 \AA\ UV LFs do not tell us anything about the ionizing escape fraction of these galaxies.  Fortunately, we have additional probes of the first billion years, which directly measure the timing of the EoR.  If high-redshift galaxies are responsible for driving the EoR, as seems highly likely, perhaps combining LF observations with EoR observations will allow us to constrain $f_{\rm esc}$.  Indeed similar approaches combining LFs and EoR measurements have been used by several studies to constrain the escape fraction and its redshift evolution \citep[e.g.][]{Haardt&Madau2012,Kuhlen2012,Mitra2013,Robertson2013,Price2016}.

In this section we expand our model parameter space to include $f_{\rm esc, 10}$ and $\alpha_{\rm esc}$, and include two additional observational data sets in our MCMC.  We use the EoR constraints which are the least model dependent: (i) the electron scattering optical depth to the cosmic microwave background (CMB) \citep{Planck2016}; and (ii) the dark fraction of pixels in QSO spectra \citep{McGreer2015}.  For (i) we use the latest estimate of the optical depth  $\tau_{\rm e} = 0.058 \pm 0.012 (1\sigma)$ from \cite{Planck2016}. For (ii) we use the upper limit from \cite{McGreer2015} $x_{\ion{H}{I}} < 0.06 + 0.05 (1\sigma)$ at $z=5.9$.  The reionization history from each sample in the chain is compared against these observations, and the corresponding $\chi^2$ likelihoods are multiplied together with the LF likelihoods.

The resulting parameter constraints are shown with the blue dashed curves in the triangle plot of Fig. \ref{fig:contour}, and summarized in the second row of table \ref{Table:recovered_parameters}.  The additional EoR observations do not improve constraints on the four star formation model parameters studied in the previous section; these are determined almost entirely by LF observations. From the marginalized 1D PDFs, the normalization parameter, ${\rm log_{10}}(f_{\rm esc, 10})$ is constrained to a $1\sigma$ percentage error of $\sim 40$ per cent. Although this is two times larger than we can obtain with 21-cm signal in the next section, it is the best constraint currently available. It shows a mild degeneracy with the star formation parameters, $f_{\ast, 10}$ and $t_\ast$, which are mostly constrained with LF observations.  The scaling of the escape fraction with halo mass, $\alpha_{\rm esc}$, is entirely unconstrained, as evidenced by the flat PDFs, which are very similar to our flat priors.

In Appendix \ref{app:Emissivity} we also consider measurements of the ionizing emissivity at lower redshifts.  These can be used as upper limits for the galaxy ionizing emission, since at those redshifts ($ z\lsim 5$), the contribution from AGN could be non-negligible.  We show that current measurements, although not adding constraining power to most of the galaxy formation parameters, do reduce the $1\sigma$ error for the escape fraction down to $\sim 25$ per cent.  This is however a very model-dependent measurement, and so we leave it out of our fiducial data sets.

  \subsubsection{Corresponding luminosity functions inferred from the data}

  One powerful benefit of our model is that it allows us to infer LFs down to the faint galaxies and high redshifts, inaccessible by current observations.  These LFs, corresponding to the astrophysical parameter constraints, are shown in the top right corner of Fig. \ref{fig:contour}, with the blue hatched region corresponding to the 95\% C.L..  From the panels we see explicitly that the faint end, $M_{\rm UV} \gsim -14$ is poorly constrained, consistent with our broad marginalized limits on $M_{\rm turn}$.  However, due to the tight constraints on $\alpha_\ast$ and the ratio $f_{\ast, 10} / t_\ast$, current observations allow us to place very tight constraints on $M_{\rm UV} \lsim -14$ galaxies.  This is true even at very high redshifts, e.g. $z\sim15$, where we have {\it no} observations currently.  These Bayesian constraints are a by-product of our model assumption that there is a redshift independent relation between the stellar mass and halo mass, and that the stellar mass is built-up with a rate which scales inversely with the Hubble time.  These assumptions allow us to constrain high-$z$ LFs even with current observations.

\subsection{Using only 21-cm signal}\label{sec:Constarint-21cm}

We now turn to the constraining power of the 21-cm signal. In \citet{Greig2017, Greig2018} we showed that mock 21-cm power spectra alone could constrain a simpler parametrization of galaxy properties.\footnote{ \cite{Mirocha2015} presented parameter constraints from measurements of the inflection points of {\it the global 21-cm signal}, using an idealized instrument. Their results are directly comparable to constraints from the power spectra in \cite{Greig2017}. Depending on the fiducial model, they recover four astrophysical parameters to roughly $\sim 10$ per cent precision, while \cite{Greig2017} recover five comparable astrophysical parameters to of order $\sim$1 percent precision, using 21-cm power spectra. This improvement in precision despite a larger model parameter space is due to the power spectrum encoding information about the structure of the signal which breaks degeneracies in the timing obtained from the global signal (e.g. Fig. 2 in \citealp{21CMMC}), even when realistic noise and foreground avoidance is assumed for the interferometer.} Our model here is more sophisticated/flexible with more free parameters, and so we expect constraints to be weaker.

Unlike in the previous section, here we have no current observations to use for our MCMC.  We therefore use a mock 1000h 21-cm observation, generated from different cosmological initial conditions, as described in \S \ref{sec:mock}.  This mock observation is created using the fiducial parameters shown in Table 1, and denoted with the vertical and horizontal dotted lines in our corner plot.  Although these fiducial choices are consistent at 1$\,\sigma$ with those recovered from actual LF observations, they do not correspond to the ML values.  As such, the LF data and 21-cm data do not converge to a single set of parameters, and thus we slightly underestimate the potential of combining the two measurements, quantified in the following section.  This is a reasonably conservative assumption, as there could be unknown systematics presented in either observation which could pull the posterior toward different  values.

In Fig.~\ref{fig:contour} (pink solid lines) we show the $95$ percentiles for each of the eight free parameters along with the 1D marginalized PDFs. In the top right of the figure, we show the recovered median values of the IGM neutral fraction, ${\bar x}_{\ion{H}{I}}$, with $95$ per cent confidence levels.
It is clear that we recover the parameters used in the mock observation.  Specifically, the recovered $68\%$ confidence intervals are [${\rm log_{10}}(f_{\ast,10})$, $\alpha_{\ast}$, ${\rm log_{10}}(f_{\rm esc,10})$, $\alpha_{\rm esc}$, ${\rm log_{10}}(M_{\rm turn})$, $t_{\ast}$, ${\rm log_{10}}(L_{\rm X<2\,keV}/{\rm SFR})$, $E_0$] $=$ ($15$, $71$, $23$, $63$, $3$, $34$, $0.2$, $8$) per cent.

From the marginalized 1D PDFs, we see that the 21-cm signal is not very sensitive to $\alpha_\ast$ alone.  The LF observations, discussed in the previous section and shown with the green curve, are much more powerful at constraining this scaling of the SFR with halo mass.

Other parameters are recovered at either comparable confidence as using LF alone, or with improved confidence.  For example, the turnover halo mass, ${\rm log_{10}}(M_{\rm turn})$, is constrained to a $1\,\sigma$ percentage error of $3$, indicating that the 21-cm signal can inform us on the turnover scale which is not captured by LF observations. However, it does show a degeneracy with $\alpha_{\ast}$.  Both $M_{\rm turn}$ and $\alpha_\ast$ help in determining which DM mass scale hosts the dominant population of star forming galaxies which drive the 21-cm signal: increasing either shifts the population towards higher mass halos, and visa versa.  As the large-scale 21-cm power is sensitive to the bias of the average (unseen) source population \citep[e.g][]{McQuinn2007,McQuinn&D'Aloisio2018}, our mock observation constrains a combination of these two parameters.

There is also a degeneracy between the normalization and halo mass scalings of the escape fraction and the star formation efficiency, as evidenced by the $f_{\ast, 10}$ vs $f_{\rm esc, 10}$ and $\alpha_\ast$ vs $\alpha_{\rm esc}$ panels.  This is understandable as the epoch of reionization, which is at the lowest redshifts for which the telescopes are most sensitive (see Appendix~\ref{app:Mock}), only depends on the product of $f_\ast$ and $f_{\rm esc}$ (c.f. eq. \ref{eq:nion}).  On the other hand, the EoH and WF coupling epochs only depend on $f_\ast$, ameliorating the degeneracy.

We also note a degeneracy in $f_{\ast,10}$ vs $t_{\ast}$, although it is smaller than when using LFs only. In contrast with LFs that are only sensitive to the instantaneous SFR, the 21-cm EoR signal more strongly depends on the {\it cumulative} SFR (i.e. the stellar mass), since the average IGM recombination timescale is longer than the duration of the EoR; thus once a cosmic IGM patch is ionized, it generally stays ionized. Nevertheless, since the comoving specific emissivity, which is used for the EoH and WF coupling epochs [equation~(\ref{eq:specific_emissivity})] is still proportional to $f_{\ast,10}/t_{\ast}$, the degeneracy is not completely broken.

Finally, we note that the X-ray properties of the first galaxies, inaccessible with UV LFs, are very strongly constrained with the 21-cm signal. In particular, the soft-band X-ray luminosity per unit SFR can be constrained at the level of $\sim$0.1 percent while the minimum X-ray energy escaping the galaxies (which is related to the typical ISM column density) can be constrained at $\sim 1$ -- 10 percent, as seen from the 1D marginalized PDFs. \footnote{This statement is true for our fiducial parameter set used to calculate the mock observation.  As quantified by \citet{Gillet2018}, if the ISM attenuation of early galaxies is much larger than we expect, such that only hard X-rays escape to heat the IGM (see also \citealp{Mesinger2013,Fialkov&Barkana2014}), $E_0$ will not be recovered.  This is due to the strong dependence of the absorption cross section to photon energy, making the EoH insensitive to hard X-rays.}

\subsection{Using both LFs and the 21-cm signal}\label{sec:combined_constraints}

Finally, we show parameter constraints if both the LF observations and the mock 21cm observations are used when computing the likelihood.
The resulting marginalized distributions are shown as shaded regions in the triangle plot of Fig. \ref{fig:contour}, and the corresponding 2 $\sigma$ constraints on the EoR history are shown with the gray lines in the inset of the figure.

As expected, all of the constraints are either similar to or improved when combining both data sets.  As noted earlier, these results are conservative in that the ML values for the LF only MCMC are not used to create the mock 21-cm signal; thus the two data sets pull the posterior towards slightly different values (c.f. the $f_{\ast, 10}$ 1D PDFs), crudely mimicking the impact on unknown systematics.

In general, the two data sets are fairly complementary, with 21-cm providing the bulk of the constraining power. $M_{\rm turn}$, $t_\ast$, $f_{\rm esc, 10}$, $\alpha_{\rm esc}$, $L_{\rm X<2 keV}/{\rm SFR}$, and $E_0$ are determined almost entirely by the 21-cm signal. $\alpha_\ast$ is determined almost entirely by the LFs, while $f_{\ast, 10}$ is constrained by both data sets to a comparable degree. This is evident also in the corresponding LF constraints, in which the bright end is constrained by both data sets, while the faint end is more strongly constrained by 21-cm.  Moreover, the $f_{\ast, 10}$ - $t_\ast$ degeneracy is significantly mitigated by combining the data sets.

From the middle-right panel, we see that the EoR history is entirely constrained by 21-cm.  Although knowing the EoR history is less remarkable than knowing the various galaxy properties in the triangle plot, it enables 21-cm observations to tightly constrain $\tau_e$: an important systematic for CMB studies \citep[e.g.][]{Liu2016}.

In summary, the $1\,\sigma$ percentage errors on our parameters from the combined data sets are [${\rm log_{10}}(f_{\ast,10})$, $\alpha_{\ast}$, ${\rm log_{10}}(f_{\rm esc,10})$, $\alpha_{\rm esc}$, ${\rm log_{10}}(M_{\rm turn})$, $t_{\ast}$, ${\rm log_{10}}(L_{\rm X<2\,keV}/{\rm SFR})$, $E_0$] $=$ ($12$, $13$, $15$, $33$, $2.4$, $33$, $0.14$, $6$) per cent.  \footnote{We note that the tightest constraints we obtain are on  $L_{\rm X<2 keV}/{\rm SFR}$, constrained to $\sim 1.4$ per cent.  However, this is only strictly true in the context of our model.  For example, if one allows $L_{\rm X<2 keV}/{\rm SFR}$ to vary with host halo mass or time, it will be less tightly constrained due to the additional free parameter(s). Nevertheless, the power of our fully Bayesian framework is that when we have an actual observation, we can easily test whether or not the data prefer a more complicated model, using the evidence to perform model selection.}

%
%
\section{Conclusion}\label{sec:conclusion}

In the near future we will detect the 3D structure of the cosmic 21-cm signal.  This signal promises to be a treasure trove of physics, informing us on the properties of the otherwise unseen population of galaxies driving the EoR and CD.

Here we develop an expanded, flexible model for galaxy formation, implementing it in the 21-cm modeling code {\cmfst}.  In particular: (i) we allow both the stellar mass and the ionizing escape fraction to be a function of the mass of the host halo; (ii) we implement a duty cycle which suppresses star formation inside low mass halos; (iii) we directly incorporate sub-grid recombinations based on the local density and ionization history.

Using a Monte Carlo Markov Chain sampler of 3D simulations, {\cmmc}, we constrain the eight free parameters of our galaxy model using: (i) current observations of high-$z$ luminosity functions (LFs); (ii) mock 21-cm power spectra as measured by a 1000h integration with HERA; (iii) and a combination of (i) and (ii).

Using only UV LFs allows us to constrain the scaling of the star formation efficiency with halo mass, and the ratio of $f_{\ast,10} / t_\ast$.  Folding-in EoR observations allows us to additionally weakly constrain the normalization of the ionizing escape fraction, $f_{\rm esc, 10}$, but not its dependence on the halo mass.

Including the mock 21-cm power spectra when performing inference allows us to mitigate these degeneracies, constraining even the ionizing escape fraction and two additional X-ray properties: (i) the soft band X-ray luminosity per unit star formation, and (ii) the minimum X-ray energy escaping the galaxies (analogous to the typical ISM column density).  The halo mass scaling, and to a lesser extent the normalization, of the stellar mass is mostly constrained by the LFs. The remaining parameters are mostly constrained by the 21-cm power spectra.  Combining the two parameter sets, we recover all of the parameters at the level of $\sim10$\% or better, with only mild degeneracies remaining.

Our flexible framework makes it easy to tie galaxy observations to the corresponding 21-cm signal. Moreover, 21-cm forecasts can be made from more detailed semi-analytic models of galaxy formation, by casting them into our framework. These improvements to our modeling and inference codes are made publicly available at {\cmfst} (\url{https://github.com/andreimesinger/21cmFAST}) and {\cmmc} (\url{https://github.com/BradGreig/21CMMC}).


\section*{Acknowledgements}

We thank R. Bouwens, S. Finkelstein and R. Livermore for their UV LF data, and associated insightful discussions. This work was supported by the European Research Council (ERC) under the European Union's Horizon 2020 research and innovation programme (grant agreement No 638809 -- AIDA -- PI: Mesinger). The results presented here reflect the authors' views; the ERC is not responsible for their use.  Parts of this research were supported by the Australian Research Council Centre of Excellence for All Sky Astrophysics in 3 Dimensions (ASTRO 3D), through project number CE170100013.  We acknowledge support from INAF under PRIN SKA/CTA FORECaST.
The simulations shown in Appendix \ref{app:LFs} were performed as part of the PRACE tier-0 grant GAFFER (project No 2016163945). We acknowledge PRACE for awarding us access to Curie at GENCI@CEA, France.




\bibliographystyle{mnras}
\bibliography{ref.bib} 

\begin{thebibliography}{}
\makeatletter
\relax
\def\mn@urlcharsother{\let\do\@makeother \do\$\do\&\do\#\do\^\do\_\do\%\do\~}
\def\mn@doi{\begingroup\mn@urlcharsother \@ifnextchar [ {\mn@doi@}
  {\mn@doi@[]}}
\def\mn@doi@[#1]#2{\def\@tempa{#1}\ifx\@tempa\@empty \href
  {http://dx.doi.org/#2} {doi:#2}\else \href {http://dx.doi.org/#2} {#1}\fi
  \endgroup}
\def\mn@eprint#1#2{\mn@eprint@#1:#2::\@nil}
\def\mn@eprint@arXiv#1{\href {http://arxiv.org/abs/#1} {{\tt arXiv:#1}}}
\def\mn@eprint@dblp#1{\href {http://dblp.uni-trier.de/rec/bibtex/#1.xml}
  {dblp:#1}}
\def\mn@eprint@#1:#2:#3:#4\@nil{\def\@tempa {#1}\def\@tempb {#2}\def\@tempc
  {#3}\ifx \@tempc \@empty \let \@tempc \@tempb \let \@tempb \@tempa \fi \ifx
  \@tempb \@empty \def\@tempb {arXiv}\fi \@ifundefined
  {mn@eprint@\@tempb}{\@tempb:\@tempc}{\expandafter \expandafter \csname
  mn@eprint@\@tempb\endcsname \expandafter{\@tempc}}}

\bibitem[\protect\citeauthoryear{{Akeret}, {Seehars}, {Amara}, {Refregier}  \&
  {Csillaghy}}{{Akeret} et~al.}{2013}]{Akeret2013}
{Akeret} J.,  {Seehars} S.,  {Amara} A.,  {Refregier} A.,   {Csillaghy} A.,
  2013, \mn@doi [Astronomy and Computing] {10.1016/j.ascom.2013.06.003}, \href
  {http://adsabs.harvard.edu/abs/2013A%26C.....2...27A} {2, 27}

\bibitem[\protect\citeauthoryear{{Alvarez} \& {Abel}}{{Alvarez} \&
  {Abel}}{2012}]{Alvarez&Abel2012}
{Alvarez} M.~A.,  {Abel} T.,  2012, \mn@doi [\apj]
  {10.1088/0004-637X/747/2/126}, \href
  {http://adsabs.harvard.edu/abs/2012ApJ...747..126A} {747, 126}

\bibitem[\protect\citeauthoryear{{Atek}, {Richard}, {Kneib}  \&
  {Schaerer}}{{Atek} et~al.}{2018}]{Atek2018}
{Atek} H.,  {Richard} J.,  {Kneib} J.-P.,   {Schaerer} D.,  2018, preprint,
  \href {http://adsabs.harvard.edu/abs/2018arXiv180309747A} {} (\mn@eprint
  {arXiv} {1803.09747})

\bibitem[\protect\citeauthoryear{{Aubert}, {Deparis}  \& {Ocvirk}}{{Aubert}
  et~al.}{2015}]{Aubert2015}
{Aubert} D.,  {Deparis} N.,   {Ocvirk} P.,  2015, \mn@doi [\mnras]
  {10.1093/mnras/stv1896}, \href
  {http://cdsads.u-strasbg.fr/abs/2015MNRAS.454.1012A} {454, 1012}

\bibitem[\protect\citeauthoryear{{Baek} \& {Ferrara}}{{Baek} \&
  {Ferrara}}{2013}]{Baek&Ferrara2013}
{Baek} S.,  {Ferrara} A.,  2013, \mn@doi [\mnras] {10.1093/mnrasl/slt023},
  \href {http://adsabs.harvard.edu/abs/2013MNRAS.432L...6B} {432, L6}

\bibitem[\protect\citeauthoryear{{Barkana} \& {Loeb}}{{Barkana} \&
  {Loeb}}{2001}]{Barkana&Loeb2001}
{Barkana} R.,  {Loeb} A.,  2001, \mn@doi [\physrep]
  {10.1016/S0370-1573(01)00019-9}, \href
  {http://adsabs.harvard.edu/abs/2001PhR...349..125B} {349, 125}

\bibitem[\protect\citeauthoryear{{Barkana} \& {Loeb}}{{Barkana} \&
  {Loeb}}{2004}]{Barkana&Loeb2004}
{Barkana} R.,  {Loeb} A.,  2004, \mn@doi [\apj] {10.1086/421079}, \href
  {http://adsabs.harvard.edu/abs/2004ApJ...609..474B} {609, 474}

\bibitem[\protect\citeauthoryear{{Barkana} \& {Loeb}}{{Barkana} \&
  {Loeb}}{2005}]{Barkana&Loeb2005}
{Barkana} R.,  {Loeb} A.,  2005, \mn@doi [\apj] {10.1086/429954}, \href
  {http://adsabs.harvard.edu/abs/2005ApJ...626....1B} {626, 1}

\bibitem[\protect\citeauthoryear{{Barkana} \& {Loeb}}{{Barkana} \&
  {Loeb}}{2008}]{Barkana&Loeb2008}
{Barkana} R.,  {Loeb} A.,  2008, \mn@doi [\mnras]
  {10.1111/j.1365-2966.2007.12729.x}, \href
  {http://adsabs.harvard.edu/abs/2008MNRAS.384.1069B} {384, 1069}

\bibitem[\protect\citeauthoryear{{Basu-Zych} et~al.,}{{Basu-Zych}
  et~al.}{2013}]{Basu-Zych2013}
{Basu-Zych} A.~R.,  et~al., 2013, \mn@doi [\apj] {10.1088/0004-637X/762/1/45},
  \href {http://adsabs.harvard.edu/abs/2013ApJ...762...45B} {762, 45}

\bibitem[\protect\citeauthoryear{{Beardsley}, {Morales}, {Lidz}, {Malloy}  \&
  {Sutter}}{{Beardsley} et~al.}{2015}]{Beardsley2015}
{Beardsley} A.~P.,  {Morales} M.~F.,  {Lidz} A.,  {Malloy} M.,   {Sutter}
  P.~M.,  2015, \mn@doi [\apj] {10.1088/0004-637X/800/2/128}, \href
  {http://adsabs.harvard.edu/abs/2015ApJ...800..128B} {800, 128}

\bibitem[\protect\citeauthoryear{{Becker}, {Bolton}, {Madau}, {Pettini},
  {Ryan-Weber}  \& {Venemans}}{{Becker} et~al.}{2015}]{Becker2015}
{Becker} G.~D.,  {Bolton} J.~S.,  {Madau} P.,  {Pettini} M.,  {Ryan-Weber}
  E.~V.,   {Venemans} B.~P.,  2015, \mn@doi [\mnras] {10.1093/mnras/stu2646},
  \href {http://adsabs.harvard.edu/abs/2015MNRAS.447.3402B} {447, 3402}

\bibitem[\protect\citeauthoryear{Behroozi \& Silk}{Behroozi \&
  Silk}{2015}]{Behroozi2015}
Behroozi P.~S.,  Silk J.,  2015, \mn@doi [Astrophysical Journal]
  {10.1088/0004-637X/799/1/32}, 799

\bibitem[\protect\citeauthoryear{Bouwens et~al.,}{Bouwens
  et~al.}{2012}]{Bouwens2012}
Bouwens R.~J.,  et~al., 2012, \mn@doi [Astrophysical Journal]
  {10.1088/0004-637X/754/2/83}, 754

\bibitem[\protect\citeauthoryear{Bouwens et~al.,}{Bouwens
  et~al.}{2015a}]{Bouwens2015_LF4-10}
Bouwens R.~J.,  et~al., 2015a, \mn@doi [Astrophysical Journal]
  {10.1088/0004-637X/803/1/34}, 803, 1

\bibitem[\protect\citeauthoryear{{Bouwens}, {Illingworth}, {Oesch}, {Caruana},
  {Holwerda}, {Smit}  \& {Wilkins}}{{Bouwens} et~al.}{2015b}]{Bouwens2015}
{Bouwens} R.~J.,  {Illingworth} G.~D.,  {Oesch} P.~A.,  {Caruana} J.,
  {Holwerda} B.,  {Smit} R.,   {Wilkins} S.,  2015b, \mn@doi [\apj]
  {10.1088/0004-637X/811/2/140}, \href
  {http://adsabs.harvard.edu/abs/2015ApJ...811..140B} {811, 140}

\bibitem[\protect\citeauthoryear{Bouwens, Oesch, Illingworth, Ellis  \&
  Stefanon}{Bouwens et~al.}{2016}]{Bouwens2016_LF6}
Bouwens R.~J.,  Oesch P.~A.,  Illingworth G.~D.,  Ellis R.~S.,   Stefanon M.,
  2016, \mn@doi [The Astrophysical Journal] {10.3847/1538-4357/aa70a4}, 843,
  129

\bibitem[\protect\citeauthoryear{{Bowman} \& {Rogers}}{{Bowman} \&
  {Rogers}}{2010}]{Bowman&Rogers2010}
{Bowman} J.~D.,  {Rogers} A.~E.~E.,  2010, \mn@doi [\nat]
  {10.1038/nature09601}, \href
  {http://adsabs.harvard.edu/abs/2010Natur.468..796B} {468, 796}

\bibitem[\protect\citeauthoryear{{Bowman} et~al.,}{{Bowman}
  et~al.}{2013}]{Bowman2013}
{Bowman} J.~D.,  et~al., 2013, \mn@doi [\pasa] {10.1017/pas.2013.009}, \href
  {http://adsabs.harvard.edu/abs/2013PASA...30...31B} {30, e031}

\bibitem[\protect\citeauthoryear{{Bowman}, {Rogers}, {Monsalve}, {Mozdzen}  \&
  {Mahesh}}{{Bowman} et~al.}{2018}]{Bowman2018}
{Bowman} J.~D.,  {Rogers} A.~E.~E.,  {Monsalve} R.~A.,  {Mozdzen} T.~J.,
  {Mahesh} N.,  2018, \mn@doi [\nat] {10.1038/nature25792}, \href
  {http://adsabs.harvard.edu/abs/2018Natur.555...67B} {555, 67}

\bibitem[\protect\citeauthoryear{{Brorby}, {Kaaret}, {Prestwich}  \&
  {Mirabel}}{{Brorby} et~al.}{2016}]{Brorby2016}
{Brorby} M.,  {Kaaret} P.,  {Prestwich} A.,   {Mirabel} I.~F.,  2016, \mn@doi
  [\mnras] {10.1093/mnras/stw284}, \href
  {http://adsabs.harvard.edu/abs/2016MNRAS.457.4081B} {457, 4081}

\bibitem[\protect\citeauthoryear{Capak et~al.,}{Capak et~al.}{2015}]{Capak2015}
Capak P.~L.,  et~al., 2015, \mn@doi [Nature] {10.1038/nature14500}, 522, 455

\bibitem[\protect\citeauthoryear{{Chardin}, {Haehnelt}, {Aubert}  \&
  {Puchwein}}{{Chardin} et~al.}{2015}]{Chardin2015}
{Chardin} J.,  {Haehnelt} M.~G.,  {Aubert} D.,   {Puchwein} E.,  2015, \mn@doi
  [\mnras] {10.1093/mnras/stv1786}, \href
  {http://adsabs.harvard.edu/abs/2015MNRAS.453.2943C} {453, 2943}

\bibitem[\protect\citeauthoryear{{Choudhury} \& {Ferrara}}{{Choudhury} \&
  {Ferrara}}{2006}]{Choudhury&Ferrara2006}
{Choudhury} T.~R.,  {Ferrara} A.,  2006, \mn@doi [\mnras]
  {10.1111/j.1745-3933.2006.00207.x}, \href
  {http://adsabs.harvard.edu/abs/2006MNRAS.371L..55C} {371, L55}

\bibitem[\protect\citeauthoryear{{Ciardi}, {Scannapieco}, {Stoehr}, {Ferrara},
  {Iliev}  \& {Shapiro}}{{Ciardi} et~al.}{2006}]{Ciardi2006}
{Ciardi} B.,  {Scannapieco} E.,  {Stoehr} F.,  {Ferrara} A.,  {Iliev} I.~T.,
  {Shapiro} P.~R.,  2006, \mn@doi [\mnras] {10.1111/j.1365-2966.2005.09908.x},
  \href {http://adsabs.harvard.edu/abs/2006MNRAS.366..689C} {366, 689}

\bibitem[\protect\citeauthoryear{{D'Aloisio}, {McQuinn}, {Maupin}, {Davies},
  {Trac}, {Fuller}  \& {Upton Sanderbeck}}{{D'Aloisio}
  et~al.}{2018a}]{D'Aloisio2018b}
{D'Aloisio} A.,  {McQuinn} M.,  {Maupin} O.,  {Davies} F.~B.,  {Trac} H.,
  {Fuller} S.,   {Upton Sanderbeck} P.~R.,  2018a, preprint, \href
  {http://adsabs.harvard.edu/abs/2018arXiv180709282D} {} (\mn@eprint {arXiv}
  {1807.09282})

\bibitem[\protect\citeauthoryear{{D'Aloisio}, {McQuinn}, {Davies}  \&
  {Furlanetto}}{{D'Aloisio} et~al.}{2018b}]{D'Aloisio2018}
{D'Aloisio} A.,  {McQuinn} M.,  {Davies} F.~B.,   {Furlanetto} S.~R.,  2018b,
  \mn@doi [\mnras] {10.1093/mnras/stx2341}, \href
  {http://adsabs.harvard.edu/abs/2018MNRAS.473..560D} {473, 560}

\bibitem[\protect\citeauthoryear{{Das}, {Mesinger}, {Pallottini}, {Ferrara}  \&
  {Wise}}{{Das} et~al.}{2017}]{Das2017}
{Das} A.,  {Mesinger} A.,  {Pallottini} A.,  {Ferrara} A.,   {Wise} J.~H.,
  2017, \mn@doi [\mnras] {10.1093/mnras/stx943}, \href
  {http://adsabs.harvard.edu/abs/2017MNRAS.469.1166D} {469, 1166}

\bibitem[\protect\citeauthoryear{Dayal, Ferrara, Dunlop  \& Pacucci}{Dayal
  et~al.}{2014}]{Dayal2014}
Dayal P.,  Ferrara A.,  Dunlop J.~S.,   Pacucci F.,  2014, \mn@doi [Monthly
  Notices of the Royal Astronomical Society] {10.1093/mnras/stu1848}, 445, 2545

\bibitem[\protect\citeauthoryear{{DeBoer} et~al.,}{{DeBoer}
  et~al.}{2017}]{DeBoer2017}
{DeBoer} D.~R.,  et~al., 2017, \mn@doi [\pasp]
  {10.1088/1538-3873/129/974/045001}, \href
  {http://adsabs.harvard.edu/abs/2017PASP..129d5001D} {129, 045001}

\bibitem[\protect\citeauthoryear{{Eide}, {Graziani}, {Ciardi}, {Feng},
  {Kakiichi}  \& {Di Matteo}}{{Eide} et~al.}{2018}]{Eide2018}
{Eide} M.~B.,  {Graziani} L.,  {Ciardi} B.,  {Feng} Y.,  {Kakiichi} K.,   {Di
  Matteo} T.,  2018, \mn@doi [\mnras] {10.1093/mnras/sty272}, \href
  {http://adsabs.harvard.edu/abs/2018MNRAS.476.1174E} {476, 1174}

\bibitem[\protect\citeauthoryear{{Ferrara} \& {Loeb}}{{Ferrara} \&
  {Loeb}}{2013}]{Ferrara&Loeb2013}
{Ferrara} A.,  {Loeb} A.,  2013, \mn@doi [\mnras] {10.1093/mnras/stt381}, \href
  {http://adsabs.harvard.edu/abs/2013MNRAS.431.2826F} {431, 2826}

\bibitem[\protect\citeauthoryear{{Fialkov} \& {Barkana}}{{Fialkov} \&
  {Barkana}}{2014}]{Fialkov&Barkana2014}
{Fialkov} A.,  {Barkana} R.,  2014, \mn@doi [\mnras] {10.1093/mnras/stu1744},
  \href {http://adsabs.harvard.edu/abs/2014MNRAS.445..213F} {445, 213}

\bibitem[\protect\citeauthoryear{{Fialkov}, {Cohen}, {Barkana}  \&
  {Silk}}{{Fialkov} et~al.}{2017}]{Fialkov2017}
{Fialkov} A.,  {Cohen} A.,  {Barkana} R.,   {Silk} J.,  2017, \mn@doi [\mnras]
  {10.1093/mnras/stw2540}, \href
  {http://adsabs.harvard.edu/abs/2017MNRAS.464.3498F} {464, 3498}

\bibitem[\protect\citeauthoryear{{Field}}{{Field}}{1958}]{Field1958}
{Field} G.~B.,  1958, \mn@doi [Proceedings of the IRE]
  {10.1109/JRPROC.1958.286741}, \href
  {http://adsabs.harvard.edu/abs/1958PIRE...46..240F} {46, 240}

\bibitem[\protect\citeauthoryear{{Field}}{{Field}}{1959}]{Field1959}
{Field} G.~B.,  1959, \mn@doi [\apj] {10.1086/146653}, \href
  {http://adsabs.harvard.edu/abs/1959ApJ...129..536F} {129, 536}

\bibitem[\protect\citeauthoryear{{Finlator}, {Oh}, {{\"O}zel}  \&
  {Dav{\'e}}}{{Finlator} et~al.}{2012}]{Finlator2012}
{Finlator} K.,  {Oh} S.~P.,  {{\"O}zel} F.,   {Dav{\'e}} R.,  2012, \mn@doi
  [\mnras] {10.1111/j.1365-2966.2012.22114.x}, \href
  {http://adsabs.harvard.edu/abs/2012MNRAS.427.2464F} {427, 2464}

\bibitem[\protect\citeauthoryear{{Foreman-Mackey}, {Hogg}, {Lang}  \&
  {Goodman}}{{Foreman-Mackey} et~al.}{2013}]{Foreman-Mackey2013}
{Foreman-Mackey} D.,  {Hogg} D.~W.,  {Lang} D.,   {Goodman} J.,  2013, \mn@doi
  [\pasp] {10.1086/670067}, \href
  {http://adsabs.harvard.edu/abs/2013PASP..125..306F} {125, 306}

\bibitem[\protect\citeauthoryear{{Fragos} et~al.,}{{Fragos}
  et~al.}{2013}]{Fragos2013}
{Fragos} T.,  et~al., 2013, \mn@doi [\apj] {10.1088/0004-637X/764/1/41}, \href
  {http://adsabs.harvard.edu/abs/2013ApJ...764...41F} {764, 41}

\bibitem[\protect\citeauthoryear{{Furlanetto}}{{Furlanetto}}{2006}]{Furlanetto2006}
{Furlanetto} S.~R.,  2006, \mn@doi [\mnras] {10.1111/j.1365-2966.2006.10725.x},
  \href {http://adsabs.harvard.edu/abs/2006MNRAS.371..867F} {371, 867}

\bibitem[\protect\citeauthoryear{{Furlanetto}, {Zaldarriaga}  \&
  {Hernquist}}{{Furlanetto} et~al.}{2004}]{Furlanetto2004}
{Furlanetto} S.~R.,  {Zaldarriaga} M.,   {Hernquist} L.,  2004, \mn@doi [\apj]
  {10.1086/423025}, \href {http://adsabs.harvard.edu/abs/2004ApJ...613....1F}
  {613, 1}

\bibitem[\protect\citeauthoryear{{Furlanetto}, {Oh}  \& {Briggs}}{{Furlanetto}
  et~al.}{2006}]{FOB2006}
{Furlanetto} S.~R.,  {Oh} S.~P.,   {Briggs} F.~H.,  2006, \mn@doi [\physrep]
  {10.1016/j.physrep.2006.08.002}, \href
  {http://adsabs.harvard.edu/abs/2006PhR...433..181F} {433, 181}

\bibitem[\protect\citeauthoryear{{Gardner} et~al.,}{{Gardner}
  et~al.}{2006}]{Gardner2006}
{Gardner} J.~P.,  et~al., 2006, \mn@doi [\ssr] {10.1007/s11214-006-8315-7},
  \href {http://adsabs.harvard.edu/abs/2006SSRv..123..485G} {123, 485}

\bibitem[\protect\citeauthoryear{{Ghara}, {Choudhury}  \& {Datta}}{{Ghara}
  et~al.}{2015}]{Ghara2015}
{Ghara} R.,  {Choudhury} T.~R.,   {Datta} K.~K.,  2015, \mn@doi [\mnras]
  {10.1093/mnras/stu2512}, \href
  {http://adsabs.harvard.edu/abs/2015MNRAS.447.1806G} {447, 1806}

\bibitem[\protect\citeauthoryear{{Gillet}, {Mesinger}, {Greig}, {Liu}  \&
  {Ucci}}{{Gillet} et~al.}{2018}]{Gillet2018}
{Gillet} N.,  {Mesinger} A.,  {Greig} B.,  {Liu} A.,   {Ucci} G.,  2018,
  preprint, \href {http://adsabs.harvard.edu/abs/2018arXiv180502699G} {}
  (\mn@eprint {arXiv} {1805.02699})

\bibitem[\protect\citeauthoryear{{Giroux}, {Sutherland}  \& {Shull}}{{Giroux}
  et~al.}{1994}]{Giroux1994}
{Giroux} M.~L.,  {Sutherland} R.~S.,   {Shull} J.~M.,  1994, \mn@doi [\apjl]
  {10.1086/187603}, \href {http://adsabs.harvard.edu/abs/1994ApJ...435L..97G}
  {435, L97}

\bibitem[\protect\citeauthoryear{{Gnedin} \& {Kaurov}}{{Gnedin} \&
  {Kaurov}}{2014}]{Gnedin&Kaurov2014}
{Gnedin} N.~Y.,  {Kaurov} A.~A.,  2014, \mn@doi [\apj]
  {10.1088/0004-637X/793/1/30}, \href
  {http://adsabs.harvard.edu/abs/2014ApJ...793...30G} {793, 30}

\bibitem[\protect\citeauthoryear{{Gnedin} \& {Ostriker}}{{Gnedin} \&
  {Ostriker}}{1997}]{Gnedin&Ostriker1997}
{Gnedin} N.~Y.,  {Ostriker} J.~P.,  1997, \mn@doi [\apj] {10.1086/304548},
  \href {http://adsabs.harvard.edu/abs/1997ApJ...486..581G} {486, 581}

\bibitem[\protect\citeauthoryear{{Gnedin} \& {Shaver}}{{Gnedin} \&
  {Shaver}}{2004}]{Gnedin&Shaver2004}
{Gnedin} N.~Y.,  {Shaver} P.~A.,  2004, \mn@doi [\apj] {10.1086/420735}, \href
  {http://adsabs.harvard.edu/abs/2004ApJ...608..611G} {608, 611}

\bibitem[\protect\citeauthoryear{{Goodman} \& {Weare}}{{Goodman} \&
  {Weare}}{2010}]{Goodman2010}
{Goodman} J.,  {Weare} J.,  2010, \mn@doi [Communications in Applied
  Mathematics and Computational Science, Vol.~5, No.~1, p.~65-80, 2010]
  {10.2140/camcos.2010.5.65}, \href
  {http://adsabs.harvard.edu/abs/2010CAMCS...5...65G} {5, 65}

\bibitem[\protect\citeauthoryear{{Gorce}, {Douspis}, {Aghanim}  \&
  {Langer}}{{Gorce} et~al.}{2018}]{Gorce2018}
{Gorce} A.,  {Douspis} M.,  {Aghanim} N.,   {Langer} M.,  2018, \mn@doi [\aap]
  {10.1051/0004-6361/201629661}, \href
  {http://adsabs.harvard.edu/abs/2018A%26A...616A.113G} {616, A113}

\bibitem[\protect\citeauthoryear{Greig \& Mesinger}{Greig \&
  Mesinger}{2015}]{21CMMC}
Greig B.,  Mesinger A.,  2015, \mn@doi [Monthly Notices of the Royal
  Astronomical Society] {10.1093/mnras/stv571}, 449, 4246

\bibitem[\protect\citeauthoryear{{Greig} \& {Mesinger}}{{Greig} \&
  {Mesinger}}{2017a}]{Greig&Mesinger2017}
{Greig} B.,  {Mesinger} A.,  2017a, \mn@doi [\mnras] {10.1093/mnras/stw3026},
  \href {http://adsabs.harvard.edu/abs/2017MNRAS.465.4838G} {465, 4838}

\bibitem[\protect\citeauthoryear{Greig \& Mesinger}{Greig \&
  Mesinger}{2017b}]{Greig2017}
Greig B.,  Mesinger A.,  2017b, \mn@doi [Monthly Notices of the Royal
  Astronomical Society] {10.1093/mnras/stx2118}, 472, 2651

\bibitem[\protect\citeauthoryear{{Greig} \& {Mesinger}}{{Greig} \&
  {Mesinger}}{2018}]{Greig2018}
{Greig} B.,  {Mesinger} A.,  2018, \mn@doi [\mnras] {10.1093/mnras/sty796},
  \href {http://adsabs.harvard.edu/abs/2018MNRAS.477.3217G} {477, 3217}

\bibitem[\protect\citeauthoryear{{Haardt} \& {Madau}}{{Haardt} \&
  {Madau}}{2012}]{Haardt&Madau2012}
{Haardt} F.,  {Madau} P.,  2012, \mn@doi [\apj] {10.1088/0004-637X/746/2/125},
  \href {http://adsabs.harvard.edu/abs/2012ApJ...746..125H} {746, 125}

\bibitem[\protect\citeauthoryear{{Hassan}, {Dav{\'e}}, {Finlator}  \&
  {Santos}}{{Hassan} et~al.}{2017}]{Hassan2017}
{Hassan} S.,  {Dav{\'e}} R.,  {Finlator} K.,   {Santos} M.~G.,  2017, \mn@doi
  [\mnras] {10.1093/mnras/stx420}, \href
  {http://adsabs.harvard.edu/abs/2017MNRAS.468..122H} {468, 122}

\bibitem[\protect\citeauthoryear{{Hills}, {Kulkarni}, {Meerburg}  \&
  {Puchwein}}{{Hills} et~al.}{2018}]{Hills2018}
{Hills} R.,  {Kulkarni} G.,  {Meerburg} P.~D.,   {Puchwein} E.,  2018,
  preprint, \href {http://adsabs.harvard.edu/abs/2018arXiv180501421H} {}
  (\mn@eprint {arXiv} {1805.01421})

\bibitem[\protect\citeauthoryear{{Hogan} \& {Rees}}{{Hogan} \&
  {Rees}}{1979}]{Hogan&Rees1979}
{Hogan} C.~J.,  {Rees} M.~J.,  1979, \mn@doi [\mnras]
  {10.1093/mnras/188.4.791}, \href
  {http://adsabs.harvard.edu/abs/1979MNRAS.188..791H} {188, 791}

\bibitem[\protect\citeauthoryear{{Hui} \& {Gnedin}}{{Hui} \&
  {Gnedin}}{1997}]{Hui1997}
{Hui} L.,  {Gnedin} N.~Y.,  1997, \mn@doi [\mnras] {10.1093/mnras/292.1.27},
  \href {http://adsabs.harvard.edu/abs/1997MNRAS.292...27H} {292, 27}

\bibitem[\protect\citeauthoryear{{Hutter}}{{Hutter}}{2018}]{Hutter2018}
{Hutter} A.,  2018, \mn@doi [\mnras] {10.1093/mnras/sty683}, \href
  {http://adsabs.harvard.edu/abs/2018MNRAS.477.1549H} {477, 1549}

\bibitem[\protect\citeauthoryear{Ishigaki, Kawamata, Ouchi, Oguri, Shimasaku
  \& Ono}{Ishigaki et~al.}{2017}]{Ishigaki2018}
Ishigaki M.,  Kawamata R.,  Ouchi M.,  Oguri M.,  Shimasaku K.,   Ono Y.,
  2017, \mn@doi [The Astrophysical Journal] {10.3847/1538-4357/aaa544}, 854, 73

\bibitem[\protect\citeauthoryear{{Jensen} et~al.,}{{Jensen}
  et~al.}{2013}]{Jensen2013}
{Jensen} H.,  et~al., 2013, \mn@doi [\mnras] {10.1093/mnras/stt1341}, \href
  {http://adsabs.harvard.edu/abs/2013MNRAS.435..460J} {435, 460}

\bibitem[\protect\citeauthoryear{{Katz}, {Kimm}, {Haehnelt}, {Sijacki},
  {Rosdahl}  \& {Blaizot}}{{Katz} et~al.}{2018}]{Katz2018}
{Katz} H.,  {Kimm} T.,  {Haehnelt} M.,  {Sijacki} D.,  {Rosdahl} J.,
  {Blaizot} J.,  2018, \mn@doi [\mnras] {10.1093/mnras/sty1225}, \href
  {http://adsabs.harvard.edu/abs/2018MNRAS.478.4986K} {478, 4986}

\bibitem[\protect\citeauthoryear{{Kaurov} \& {Gnedin}}{{Kaurov} \&
  {Gnedin}}{2014}]{Kaurov2014}
{Kaurov} A.~A.,  {Gnedin} N.~Y.,  2014, \mn@doi [\apj]
  {10.1088/0004-637X/787/2/146}, \href
  {http://adsabs.harvard.edu/abs/2014ApJ...787..146K} {787, 146}

\bibitem[\protect\citeauthoryear{{Kennicutt}}{{Kennicutt}}{1998}]{Kennicutt1998}
{Kennicutt} Jr. R.~C.,  1998, \mn@doi [\araa] {10.1146/annurev.astro.36.1.189},
  \href {http://adsabs.harvard.edu/abs/1998ARA%26A..36..189K} {36, 189}

\bibitem[\protect\citeauthoryear{{Kern}, {Liu}, {Parsons}, {Mesinger}  \&
  {Greig}}{{Kern} et~al.}{2017}]{Kern2017}
{Kern} N.~S.,  {Liu} A.,  {Parsons} A.~R.,  {Mesinger} A.,   {Greig} B.,  2017,
  \mn@doi [\apj] {10.3847/1538-4357/aa8bb4}, \href
  {http://adsabs.harvard.edu/abs/2017ApJ...848...23K} {848, 23}

\bibitem[\protect\citeauthoryear{{Kimm}, {Katz}, {Haehnelt}, {Rosdahl},
  {Devriendt}  \& {Slyz}}{{Kimm} et~al.}{2017}]{Kimm2017}
{Kimm} T.,  {Katz} H.,  {Haehnelt} M.,  {Rosdahl} J.,  {Devriendt} J.,   {Slyz}
  A.,  2017, \mn@doi [\mnras] {10.1093/mnras/stx052}, \href
  {http://adsabs.harvard.edu/abs/2017MNRAS.466.4826K} {466, 4826}

\bibitem[\protect\citeauthoryear{{Koopmans} et~al.,}{{Koopmans}
  et~al.}{2015}]{Koopmans2015}
{Koopmans} L.,  et~al., 2015, Advancing Astrophysics with the Square Kilometre
  Array (AASKA14), \href {http://adsabs.harvard.edu/abs/2015aska.confE...1K}
  {p.~1}

\bibitem[\protect\citeauthoryear{Kuhlen \& Faucher-Gigu{\`{e}}re}{Kuhlen \&
  Faucher-Gigu{\`{e}}re}{2012}]{Kuhlen2012}
Kuhlen M.,  Faucher-Gigu{\`{e}}re C.~A.,  2012, \mn@doi [Monthly Notices of the
  Royal Astronomical Society] {10.1111/j.1365-2966.2012.20924.x}, 423, 862

\bibitem[\protect\citeauthoryear{{Lacey} \& {Cole}}{{Lacey} \&
  {Cole}}{1993}]{Lacey&Cole1993}
{Lacey} C.,  {Cole} S.,  1993, \mn@doi [\mnras] {10.1093/mnras/262.3.627},
  \href {http://adsabs.harvard.edu/abs/1993MNRAS.262..627L} {262, 627}

\bibitem[\protect\citeauthoryear{{Lehmer} et~al.,}{{Lehmer}
  et~al.}{2016}]{Lehmer2016}
{Lehmer} B.~D.,  et~al., 2016, \mn@doi [\apj] {10.3847/0004-637X/825/1/7},
  \href {http://adsabs.harvard.edu/abs/2016ApJ...825....7L} {825, 7}

\bibitem[\protect\citeauthoryear{{Liu}, {Pritchard}, {Allison}, {Parsons},
  {Seljak}  \& {Sherwin}}{{Liu} et~al.}{2016}]{Liu2016}
{Liu} A.,  {Pritchard} J.~R.,  {Allison} R.,  {Parsons} A.~R.,  {Seljak} U.,
  {Sherwin} B.~D.,  2016, \mn@doi [\prd] {10.1103/PhysRevD.93.043013}, \href
  {http://adsabs.harvard.edu/abs/2016PhRvD..93d3013L} {93, 043013}

\bibitem[\protect\citeauthoryear{Livermore, Finkelstein  \& Lotz}{Livermore
  et~al.}{2016}]{Livermore2017_LF}
Livermore R.~C.,  Finkelstein S.~L.,   Lotz J.~M.,  2016, \mn@doi [The
  Astrophysical Journal] {10.3847/1538-4357/835/2/113}, 835, 1

\bibitem[\protect\citeauthoryear{{Lusso}, {Worseck}, {Hennawi}, {Prochaska},
  {Vignali}, {Stern}  \& {O'Meara}}{{Lusso} et~al.}{2015}]{Lusso2015}
{Lusso} E.,  {Worseck} G.,  {Hennawi} J.~F.,  {Prochaska} J.~X.,  {Vignali} C.,
   {Stern} J.,   {O'Meara} J.~M.,  2015, \mn@doi [\mnras]
  {10.1093/mnras/stv516}, \href
  {http://adsabs.harvard.edu/abs/2015MNRAS.449.4204L} {449, 4204}

\bibitem[\protect\citeauthoryear{{Madau} \& {Fragos}}{{Madau} \&
  {Fragos}}{2017}]{Madau&Fragos2017}
{Madau} P.,  {Fragos} T.,  2017, \mn@doi [\apj] {10.3847/1538-4357/aa6af9},
  \href {http://adsabs.harvard.edu/abs/2017ApJ...840...39M} {840, 39}

\bibitem[\protect\citeauthoryear{{Madau}, {Meiksin}  \& {Rees}}{{Madau}
  et~al.}{1997}]{Madau1997}
{Madau} P.,  {Meiksin} A.,   {Rees} M.~J.,  1997, \mn@doi [\apj]
  {10.1086/303549}, \href {http://adsabs.harvard.edu/abs/1997ApJ...475..429M}
  {475, 429}

\bibitem[\protect\citeauthoryear{{Madau}, {Pozzetti}  \& {Dickinson}}{{Madau}
  et~al.}{1998}]{Madau1998}
{Madau} P.,  {Pozzetti} L.,   {Dickinson} M.,  1998, \mn@doi [\apj]
  {10.1086/305523}, \href {http://adsabs.harvard.edu/abs/1998ApJ...498..106M}
  {498, 106}

\bibitem[\protect\citeauthoryear{{Mao}, {Shapiro}, {Mellema}, {Iliev}, {Koda}
  \& {Ahn}}{{Mao} et~al.}{2012}]{Mao2012}
{Mao} Y.,  {Shapiro} P.~R.,  {Mellema} G.,  {Iliev} I.~T.,  {Koda} J.,   {Ahn}
  K.,  2012, \mn@doi [\mnras] {10.1111/j.1365-2966.2012.20471.x}, \href
  {http://adsabs.harvard.edu/abs/2012MNRAS.422..926M} {422, 926}

\bibitem[\protect\citeauthoryear{Mason, Trenti  \& Treu}{Mason
  et~al.}{2015}]{Mason2015}
Mason C.~A.,  Trenti M.,   Treu T.,  2015, \mn@doi [Proceedings of the
  International Astronomical Union] {10.1017/S1743921315009953}, 11, 33

\bibitem[\protect\citeauthoryear{{McGreer}, {Mesinger}  \&
  {D'Odorico}}{{McGreer} et~al.}{2015}]{McGreer2015}
{McGreer} I.~D.,  {Mesinger} A.,   {D'Odorico} V.,  2015, \mn@doi [\mnras]
  {10.1093/mnras/stu2449}, \href
  {http://adsabs.harvard.edu/abs/2015MNRAS.447..499M} {447, 499}

\bibitem[\protect\citeauthoryear{{McQuinn}}{{McQuinn}}{2012}]{McQuinn2012}
{McQuinn} M.,  2012, \mn@doi [\mnras] {10.1111/j.1365-2966.2012.21792.x}, \href
  {http://adsabs.harvard.edu/abs/2012MNRAS.426.1349M} {426, 1349}

\bibitem[\protect\citeauthoryear{{McQuinn} \& {D'Aloisio}}{{McQuinn} \&
  {D'Aloisio}}{2018}]{McQuinn&D'Aloisio2018}
{McQuinn} M.,  {D'Aloisio} A.,  2018, preprint, \href
  {http://adsabs.harvard.edu/abs/2018arXiv180608372M} {} (\mn@eprint {arXiv}
  {1806.08372})

\bibitem[\protect\citeauthoryear{{McQuinn} \& {O'Leary}}{{McQuinn} \&
  {O'Leary}}{2012}]{McQuinn&O'Leary2012}
{McQuinn} M.,  {O'Leary} R.~M.,  2012, \mn@doi [\apj]
  {10.1088/0004-637X/760/1/3}, \href
  {http://adsabs.harvard.edu/abs/2012ApJ...760....3M} {760, 3}

\bibitem[\protect\citeauthoryear{{McQuinn}, {Zahn}, {Zaldarriaga}, {Hernquist}
  \& {Furlanetto}}{{McQuinn} et~al.}{2006}]{McQuinn2006}
{McQuinn} M.,  {Zahn} O.,  {Zaldarriaga} M.,  {Hernquist} L.,   {Furlanetto}
  S.~R.,  2006, \mn@doi [\apj] {10.1086/505167}, \href
  {http://adsabs.harvard.edu/abs/2006ApJ...653..815M} {653, 815}

\bibitem[\protect\citeauthoryear{{McQuinn}, {Lidz}, {Zahn}, {Dutta},
  {Hernquist}  \& {Zaldarriaga}}{{McQuinn} et~al.}{2007}]{McQuinn2007}
{McQuinn} M.,  {Lidz} A.,  {Zahn} O.,  {Dutta} S.,  {Hernquist} L.,
  {Zaldarriaga} M.,  2007, \mn@doi [\mnras] {10.1111/j.1365-2966.2007.11489.x},
  \href {http://adsabs.harvard.edu/abs/2007MNRAS.377.1043M} {377, 1043}

\bibitem[\protect\citeauthoryear{{McQuinn}, {Oh}  \&
  {Faucher-Gigu{\`e}re}}{{McQuinn} et~al.}{2011}]{McQuinn2011}
{McQuinn} M.,  {Oh} S.~P.,   {Faucher-Gigu{\`e}re} C.-A.,  2011, \mn@doi [\apj]
  {10.1088/0004-637X/743/1/82}, \href
  {http://adsabs.harvard.edu/abs/2011ApJ...743...82M} {743, 82}

\bibitem[\protect\citeauthoryear{{Mellema} et~al.,}{{Mellema}
  et~al.}{2013}]{Mellema2013}
{Mellema} G.,  et~al., 2013, \mn@doi [Experimental Astronomy]
  {10.1007/s10686-013-9334-5}, \href
  {http://adsabs.harvard.edu/abs/2013ExA....36..235M} {36, 235}

\bibitem[\protect\citeauthoryear{{Mesinger} \& {Dijkstra}}{{Mesinger} \&
  {Dijkstra}}{2008}]{Mesinger2008}
{Mesinger} A.,  {Dijkstra} M.,  2008, \mn@doi [\mnras]
  {10.1111/j.1365-2966.2008.13776.x}, \href
  {http://adsabs.harvard.edu/abs/2008MNRAS.390.1071M} {390, 1071}

\bibitem[\protect\citeauthoryear{{Mesinger} \& {Furlanetto}}{{Mesinger} \&
  {Furlanetto}}{2007}]{Mesinger2007}
{Mesinger} A.,  {Furlanetto} S.,  2007, \mn@doi [\apj] {10.1086/521806}, \href
  {http://adsabs.harvard.edu/abs/2007ApJ...669..663M} {669, 663}

\bibitem[\protect\citeauthoryear{Mesinger, Furlanetto  \& Cen}{Mesinger
  et~al.}{2011}]{21cmfast}
Mesinger A.,  Furlanetto S.,   Cen R.,  2011, \mn@doi [Monthly Notices of the
  Royal Astronomical Society] {10.1111/j.1365-2966.2010.17731.x}, 411, 955

\bibitem[\protect\citeauthoryear{{Mesinger}, {Ferrara}  \&
  {Spiegel}}{{Mesinger} et~al.}{2013}]{Mesinger2013}
{Mesinger} A.,  {Ferrara} A.,   {Spiegel} D.~S.,  2013, \mn@doi [\mnras]
  {10.1093/mnras/stt198}, \href
  {http://adsabs.harvard.edu/abs/2013MNRAS.431..621M} {431, 621}

\bibitem[\protect\citeauthoryear{{Mesinger}, {Ewall-Wice}  \&
  {Hewitt}}{{Mesinger} et~al.}{2014}]{Mesinger2014}
{Mesinger} A.,  {Ewall-Wice} A.,   {Hewitt} J.,  2014, \mn@doi [\mnras]
  {10.1093/mnras/stu125}, \href
  {http://adsabs.harvard.edu/abs/2014MNRAS.439.3262M} {439, 3262}

\bibitem[\protect\citeauthoryear{{Mesinger}, {Greig}  \& {Sobacchi}}{{Mesinger}
  et~al.}{2016}]{Mesinger2016}
{Mesinger} A.,  {Greig} B.,   {Sobacchi} E.,  2016, \mn@doi [\mnras]
  {10.1093/mnras/stw831}, \href
  {http://adsabs.harvard.edu/abs/2016MNRAS.459.2342M} {459, 2342}

\bibitem[\protect\citeauthoryear{{Mineo}, {Gilfanov}  \& {Sunyaev}}{{Mineo}
  et~al.}{2012}]{Mineo2012}
{Mineo} S.,  {Gilfanov} M.,   {Sunyaev} R.,  2012, \mn@doi [\mnras]
  {10.1111/j.1365-2966.2011.19862.x}, \href
  {http://adsabs.harvard.edu/abs/2012MNRAS.419.2095M} {419, 2095}

\bibitem[\protect\citeauthoryear{{Miralda-Escud{\'e}}, {Haehnelt}  \&
  {Rees}}{{Miralda-Escud{\'e}} et~al.}{2000}]{Miralda-Escude2000}
{Miralda-Escud{\'e}} J.,  {Haehnelt} M.,   {Rees} M.~J.,  2000, \mn@doi [\apj]
  {10.1086/308330}, \href {http://adsabs.harvard.edu/abs/2000ApJ...530....1M}
  {530, 1}

\bibitem[\protect\citeauthoryear{{Mirocha} \& {Furlanetto}}{{Mirocha} \&
  {Furlanetto}}{2018}]{Mirocha&Furlanetto2018}
{Mirocha} J.,  {Furlanetto} S.~R.,  2018, preprint, \href
  {http://adsabs.harvard.edu/abs/2018arXiv180303272M} {} (\mn@eprint {arXiv}
  {1803.03272})

\bibitem[\protect\citeauthoryear{{Mirocha}, {Harker}  \& {Burns}}{{Mirocha}
  et~al.}{2015}]{Mirocha2015}
{Mirocha} J.,  {Harker} G.~J.~A.,   {Burns} J.~O.,  2015, \mn@doi [\apj]
  {10.1088/0004-637X/813/1/11}, \href
  {http://adsabs.harvard.edu/abs/2015ApJ...813...11M} {813, 11}

\bibitem[\protect\citeauthoryear{Mirocha, Furlanetto  \& Sun}{Mirocha
  et~al.}{2017}]{Mirocha2017}
Mirocha J.,  Furlanetto S.~R.,   Sun G.,  2017, \mn@doi [Monthly Notices of the
  Royal Astronomical Society] {10.1093/mnras/stw2412}, 464, 1365

\bibitem[\protect\citeauthoryear{{Mitra}, {Choudhury}  \& {Ferrara}}{{Mitra}
  et~al.}{2011}]{Mitra2011}
{Mitra} S.,  {Choudhury} T.~R.,   {Ferrara} A.,  2011, \mn@doi [\mnras]
  {10.1111/j.1365-2966.2011.18234.x}, \href
  {http://adsabs.harvard.edu/abs/2011MNRAS.413.1569M} {413, 1569}

\bibitem[\protect\citeauthoryear{{Mitra}, {Ferrara}  \& {Choudhury}}{{Mitra}
  et~al.}{2013}]{Mitra2013}
{Mitra} S.,  {Ferrara} A.,   {Choudhury} T.~R.,  2013, \mn@doi [\mnras]
  {10.1093/mnrasl/sls001}, \href
  {http://adsabs.harvard.edu/abs/2013MNRAS.428L...1M} {428, L1}

\bibitem[\protect\citeauthoryear{Mitra, Roy~Choudhury  \& Ferrara}{Mitra
  et~al.}{2015}]{Mitra2015}
Mitra S.,  Roy~Choudhury T.,   Ferrara A.,  2015, \mn@doi [Monthly Notices of
  the Royal Astronomical Society: Letters] {10.1093/mnrasl/slv134}, 454, L76

\bibitem[\protect\citeauthoryear{{Mitra}, {Choudhury}  \& {Ferrara}}{{Mitra}
  et~al.}{2018}]{Mitra2018}
{Mitra} S.,  {Choudhury} T.~R.,   {Ferrara} A.,  2018, \mn@doi [\mnras]
  {10.1093/mnras/stx2443}, \href
  {http://adsabs.harvard.edu/abs/2018MNRAS.473.1416M} {473, 1416}

\bibitem[\protect\citeauthoryear{{Mondal}, {Bharadwaj}, {Majumdar}, {Bera}  \&
  {Acharyya}}{{Mondal} et~al.}{2015}]{Mondal2015}
{Mondal} R.,  {Bharadwaj} S.,  {Majumdar} S.,  {Bera} A.,   {Acharyya} A.,
  2015, \mn@doi [\mnras] {10.1093/mnrasl/slv015}, \href
  {http://adsabs.harvard.edu/abs/2015MNRAS.449L..41M} {449, L41}

\bibitem[\protect\citeauthoryear{{Morales}}{{Morales}}{2005}]{Morales2005}
{Morales} M.~F.,  2005, \mn@doi [\apj] {10.1086/426730}, \href
  {http://adsabs.harvard.edu/abs/2005ApJ...619..678M} {619, 678}

\bibitem[\protect\citeauthoryear{{Morales} \& {Wyithe}}{{Morales} \&
  {Wyithe}}{2010}]{Morales&Wyithe2010}
{Morales} M.~F.,  {Wyithe} J.~S.~B.,  2010, \mn@doi [\araa]
  {10.1146/annurev-astro-081309-130936}, \href
  {http://adsabs.harvard.edu/abs/2010ARA%26A..48..127M} {48, 127}

\bibitem[\protect\citeauthoryear{Mutch, Geil, Poole, Angel, Duffy, Mesinger  \&
  Wyithe}{Mutch et~al.}{2016}]{Mutch2016}
Mutch S.~J.,  Geil P.~M.,  Poole G.~B.,  Angel P.~W.,  Duffy A.~R.,  Mesinger
  A.,   Wyithe J. S.~B.,  2016, \mn@doi [Monthly Notices of the Royal
  Astronomical Society] {10.1093/mnras/stw1506}, 462, 250

\bibitem[\protect\citeauthoryear{{O'Shea}, {Wise}, {Xu}  \& {Norman}}{{O'Shea}
  et~al.}{2015}]{O'Shea2015}
{O'Shea} B.~W.,  {Wise} J.~H.,  {Xu} H.,   {Norman} M.~L.,  2015, \mn@doi
  [\apjl] {10.1088/2041-8205/807/1/L12}, \href
  {http://adsabs.harvard.edu/abs/2015ApJ...807L..12O} {807, L12}

\bibitem[\protect\citeauthoryear{{Ocvirk} et~al.,}{{Ocvirk}
  et~al.}{2016}]{Ocvirk2016}
{Ocvirk} P.,  et~al., 2016, \mn@doi [\mnras] {10.1093/mnras/stw2036}, \href
  {http://adsabs.harvard.edu/abs/2016MNRAS.463.1462O} {463, 1462}

\bibitem[\protect\citeauthoryear{Oesch, Bouwens, Illingworth, Labbe  \&
  Stefanon}{Oesch et~al.}{2017}]{Oesch2018}
Oesch P.~A.,  Bouwens R.~J.,  Illingworth G.~D.,  Labbe I.,   Stefanon M.,
  2017, \mn@doi [The Astrophysical Journal] {10.3847/1538-4357/aab03f}, 855,
  105

\bibitem[\protect\citeauthoryear{{Okamoto}, {Gao}  \& {Theuns}}{{Okamoto}
  et~al.}{2008}]{Okamoto2008}
{Okamoto} T.,  {Gao} L.,   {Theuns} T.,  2008, \mn@doi [\mnras]
  {10.1111/j.1365-2966.2008.13830.x}, \href
  {http://adsabs.harvard.edu/abs/2008MNRAS.390..920O} {390, 920}

\bibitem[\protect\citeauthoryear{{Oke} \& {Gunn}}{{Oke} \&
  {Gunn}}{1983}]{Oke&Gunn1983}
{Oke} J.~B.,  {Gunn} J.~E.,  1983, \mn@doi [\apj] {10.1086/160817}, \href
  {http://adsabs.harvard.edu/abs/1983ApJ...266..713O} {266, 713}

\bibitem[\protect\citeauthoryear{{Paardekooper}, {Khochfar}  \& {Dalla
  Vecchia}}{{Paardekooper} et~al.}{2015}]{Paardekooper2015}
{Paardekooper} J.-P.,  {Khochfar} S.,   {Dalla Vecchia} C.,  2015, \mn@doi
  [\mnras] {10.1093/mnras/stv1114}, \href
  {http://adsabs.harvard.edu/abs/2015MNRAS.451.2544P} {451, 2544}

\bibitem[\protect\citeauthoryear{{Pallottini}, {Ferrara}, {Bovino}, {Vallini},
  {Gallerani}, {Maiolino}  \& {Salvadori}}{{Pallottini}
  et~al.}{2017}]{Pallottini2017}
{Pallottini} A.,  {Ferrara} A.,  {Bovino} S.,  {Vallini} L.,  {Gallerani} S.,
  {Maiolino} R.,   {Salvadori} S.,  2017, \mn@doi [\mnras]
  {10.1093/mnras/stx1792}, \href
  {http://adsabs.harvard.edu/abs/2017MNRAS.471.4128P} {471, 4128}

\bibitem[\protect\citeauthoryear{{Parsons} et~al.,}{{Parsons}
  et~al.}{2010}]{Parsons2010}
{Parsons} A.~R.,  et~al., 2010, \mn@doi [\aj] {10.1088/0004-6256/139/4/1468},
  \href {http://adsabs.harvard.edu/abs/2010AJ....139.1468P} {139, 1468}

\bibitem[\protect\citeauthoryear{{Parsons} et~al.,}{{Parsons}
  et~al.}{2014}]{Parsons2014}
{Parsons} A.~R.,  et~al., 2014, \mn@doi [\apj] {10.1088/0004-637X/788/2/106},
  \href {http://adsabs.harvard.edu/abs/2014ApJ...788..106P} {788, 106}

\bibitem[\protect\citeauthoryear{{Patra}, {Subrahmanyan}, {Raghunathan}  \&
  {Udaya Shankar}}{{Patra} et~al.}{2013}]{Patra2013}
{Patra} N.,  {Subrahmanyan} R.,  {Raghunathan} A.,   {Udaya Shankar} N.,  2013,
  \mn@doi [Experimental Astronomy] {10.1007/s10686-013-9336-3}, \href
  {http://adsabs.harvard.edu/abs/2013ExA....36..319P} {36, 319}

\bibitem[\protect\citeauthoryear{{Planck Collaboration} et~al.,}{{Planck
  Collaboration} et~al.}{2016a}]{PlanckXIII}
{Planck Collaboration} et~al., 2016a, \mn@doi [\aap]
  {10.1051/0004-6361/201525830}, \href
  {http://adsabs.harvard.edu/abs/2016A%26A...594A..13P} {594, A13}

\bibitem[\protect\citeauthoryear{{Planck Collaboration} et~al.,}{{Planck
  Collaboration} et~al.}{2016b}]{Planck2016}
{Planck Collaboration} et~al., 2016b, \mn@doi [\aap]
  {10.1051/0004-6361/201628897}, \href
  {http://adsabs.harvard.edu/abs/2016A%26A...596A.108P} {596, A108}

\bibitem[\protect\citeauthoryear{{Pober} et~al.,}{{Pober}
  et~al.}{2013}]{Pober2013}
{Pober} J.~C.,  et~al., 2013, \mn@doi [\aj] {10.1088/0004-6256/145/3/65}, \href
  {http://adsabs.harvard.edu/abs/2013AJ....145...65P} {145, 65}

\bibitem[\protect\citeauthoryear{{Pober} et~al.,}{{Pober}
  et~al.}{2014}]{Pober2014}
{Pober} J.~C.,  et~al., 2014, \mn@doi [\apj] {10.1088/0004-637X/782/2/66},
  \href {http://adsabs.harvard.edu/abs/2014ApJ...782...66P} {782, 66}

\bibitem[\protect\citeauthoryear{{Price}, {Trac}  \& {Cen}}{{Price}
  et~al.}{2016}]{Price2016}
{Price} L.~C.,  {Trac} H.,   {Cen} R.,  2016, preprint, \href
  {http://adsabs.harvard.edu/abs/2016arXiv160503970P} {} (\mn@eprint {arXiv}
  {1605.03970})

\bibitem[\protect\citeauthoryear{{Price} et~al.,}{{Price}
  et~al.}{2018}]{Price2018}
{Price} D.~C.,  et~al., 2018, \mn@doi [\mnras] {10.1093/mnras/sty1244}, \href
  {http://adsabs.harvard.edu/abs/2018MNRAS.478.4193P} {478, 4193}

\bibitem[\protect\citeauthoryear{{Pritchard} \& {Furlanetto}}{{Pritchard} \&
  {Furlanetto}}{2007}]{Pritchard&Furlanetto2007}
{Pritchard} J.~R.,  {Furlanetto} S.~R.,  2007, \mn@doi [\mnras]
  {10.1111/j.1365-2966.2007.11519.x}, \href
  {http://adsabs.harvard.edu/abs/2007MNRAS.376.1680P} {376, 1680}

\bibitem[\protect\citeauthoryear{{Pritchard} \& {Loeb}}{{Pritchard} \&
  {Loeb}}{2012}]{Pritchard&Loeb2012}
{Pritchard} J.~R.,  {Loeb} A.,  2012, \mn@doi [Reports on Progress in Physics]
  {10.1088/0034-4885/75/8/086901}, \href
  {http://adsabs.harvard.edu/abs/2012RPPh...75h6901P} {75, 086901}

\bibitem[\protect\citeauthoryear{{Rahmati}, {Pawlik}, {Rai{\v c}evi{\'c}}  \&
  {Schaye}}{{Rahmati} et~al.}{2013}]{Rahmati2013}
{Rahmati} A.,  {Pawlik} A.~H.,  {Rai{\v c}evi{\'c}} M.,   {Schaye} J.,  2013,
  \mn@doi [\mnras] {10.1093/mnras/stt066}, \href
  {http://adsabs.harvard.edu/abs/2013MNRAS.430.2427R} {430, 2427}

\bibitem[\protect\citeauthoryear{{Razoumov} \& {Sommer-Larsen}}{{Razoumov} \&
  {Sommer-Larsen}}{2010}]{Razoumov2010}
{Razoumov} A.~O.,  {Sommer-Larsen} J.,  2010, \mn@doi [\apj]
  {10.1088/0004-637X/710/2/1239}, \href
  {http://adsabs.harvard.edu/abs/2010ApJ...710.1239R} {710, 1239}

\bibitem[\protect\citeauthoryear{{Robertson} et~al.,}{{Robertson}
  et~al.}{2013}]{Robertson2013}
{Robertson} B.~E.,  et~al., 2013, \mn@doi [\apj] {10.1088/0004-637X/768/1/71},
  \href {http://adsabs.harvard.edu/abs/2013ApJ...768...71R} {768, 71}

\bibitem[\protect\citeauthoryear{{Schmit} \& {Pritchard}}{{Schmit} \&
  {Pritchard}}{2018}]{Schmit&Pritchard2018}
{Schmit} C.~J.,  {Pritchard} J.~R.,  2018, \mn@doi [\mnras]
  {10.1093/mnras/stx3292}, \href
  {http://adsabs.harvard.edu/abs/2018MNRAS.475.1213S} {475, 1213}

\bibitem[\protect\citeauthoryear{{Scoccimarro}}{{Scoccimarro}}{1998}]{Scoccimarro1998}
{Scoccimarro} R.,  1998, \mn@doi [\mnras] {10.1046/j.1365-8711.1998.01845.x},
  \href {http://adsabs.harvard.edu/abs/1998MNRAS.299.1097S} {299, 1097}

\bibitem[\protect\citeauthoryear{{Scott} \& {Rees}}{{Scott} \&
  {Rees}}{1990}]{Scott&Rees1990}
{Scott} D.,  {Rees} M.~J.,  1990, \mnras, \href
  {http://adsabs.harvard.edu/abs/1990MNRAS.247..510S} {247, 510}

\bibitem[\protect\citeauthoryear{{Shapiro}, {Giroux}  \& {Babul}}{{Shapiro}
  et~al.}{1994}]{Shapiro1994}
{Shapiro} P.~R.,  {Giroux} M.~L.,   {Babul} A.,  1994, \mn@doi [\apj]
  {10.1086/174120}, \href {http://adsabs.harvard.edu/abs/1994ApJ...427...25S}
  {427, 25}

\bibitem[\protect\citeauthoryear{{Shaver}, {Windhorst}, {Madau}  \& {de
  Bruyn}}{{Shaver} et~al.}{1999}]{Shaver1999}
{Shaver} P.~A.,  {Windhorst} R.~A.,  {Madau} P.,   {de Bruyn} A.~G.,  1999,
  \aap, \href {http://adsabs.harvard.edu/abs/1999A%26A...345..380S} {345, 380}

\bibitem[\protect\citeauthoryear{{Sheth} \& {Tormen}}{{Sheth} \&
  {Tormen}}{1999}]{S-T1999}
{Sheth} R.~K.,  {Tormen} G.,  1999, \mn@doi [\mnras]
  {10.1046/j.1365-8711.1999.02692.x}, \href
  {http://adsabs.harvard.edu/abs/1999MNRAS.308..119S} {308, 119}

\bibitem[\protect\citeauthoryear{{Sheth} \& {Tormen}}{{Sheth} \&
  {Tormen}}{2002}]{S-T2002}
{Sheth} R.~K.,  {Tormen} G.,  2002, \mn@doi [\mnras]
  {10.1046/j.1365-8711.2002.04950.x}, \href
  {http://adsabs.harvard.edu/abs/2002MNRAS.329...61S} {329, 61}

\bibitem[\protect\citeauthoryear{{Shimabukuro} \& {Semelin}}{{Shimabukuro} \&
  {Semelin}}{2017}]{Shimabukuro&Semelin2017}
{Shimabukuro} H.,  {Semelin} B.,  2017, \mn@doi [\mnras]
  {10.1093/mnras/stx734}, \href
  {http://adsabs.harvard.edu/abs/2017MNRAS.468.3869S} {468, 3869}

\bibitem[\protect\citeauthoryear{{Sobacchi} \& {Mesinger}}{{Sobacchi} \&
  {Mesinger}}{2013a}]{Sobacchi2013a}
{Sobacchi} E.,  {Mesinger} A.,  2013a, \mn@doi [\mnras]
  {10.1093/mnrasl/slt035}, \href
  {http://adsabs.harvard.edu/abs/2013MNRAS.432L..51S} {432, L51}

\bibitem[\protect\citeauthoryear{{Sobacchi} \& {Mesinger}}{{Sobacchi} \&
  {Mesinger}}{2013b}]{Sobacchi2013b}
{Sobacchi} E.,  {Mesinger} A.,  2013b, \mn@doi [\mnras] {10.1093/mnras/stt693},
  \href {http://adsabs.harvard.edu/abs/2013MNRAS.432.3340S} {432, 3340}

\bibitem[\protect\citeauthoryear{Sobacchi \& Mesinger}{Sobacchi \&
  Mesinger}{2014}]{SM14}
Sobacchi E.,  Mesinger A.,  2014, \mn@doi [Monthly Notices of the Royal
  Astronomical Society] {10.1093/mnras/stu377}, 440, 1662

\bibitem[\protect\citeauthoryear{{Sokolowski} et~al.,}{{Sokolowski}
  et~al.}{2015}]{Sokolowski2015}
{Sokolowski} M.,  et~al., 2015, \mn@doi [\pasa] {10.1017/pasa.2015.3}, \href
  {http://adsabs.harvard.edu/abs/2015PASA...32....4S} {32, e004}

\bibitem[\protect\citeauthoryear{{Somerville} \& {Kolatt}}{{Somerville} \&
  {Kolatt}}{1999}]{Somerville&Kolatt1999}
{Somerville} R.~S.,  {Kolatt} T.~S.,  1999, \mn@doi [\mnras]
  {10.1046/j.1365-8711.1999.02154.x}, \href
  {http://adsabs.harvard.edu/abs/1999MNRAS.305....1S} {305, 1}

\bibitem[\protect\citeauthoryear{{Springel} \& {Hernquist}}{{Springel} \&
  {Hernquist}}{2003}]{Springel&Hernquist2003}
{Springel} V.,  {Hernquist} L.,  2003, \mn@doi [\mnras]
  {10.1046/j.1365-8711.2003.06207.x}, \href
  {http://adsabs.harvard.edu/abs/2003MNRAS.339..312S} {339, 312}

\bibitem[\protect\citeauthoryear{Sun \& Furlanetto}{Sun \&
  Furlanetto}{2016}]{Sun&Furlanetto2016}
Sun G.,  Furlanetto S.~R.,  2016, \mn@doi [Monthly Notices of the Royal
  Astronomical Society] {10.1093/mnras/stw980}, 460, 417

\bibitem[\protect\citeauthoryear{{Tingay} et~al.,}{{Tingay}
  et~al.}{2013}]{Tingay2013}
{Tingay} S.~J.,  et~al., 2013, \mn@doi [\pasa] {10.1017/pasa.2012.007}, \href
  {http://adsabs.harvard.edu/abs/2013PASA...30....7T} {30, e007}

\bibitem[\protect\citeauthoryear{{Tozzi}, {Madau}, {Meiksin}  \&
  {Rees}}{{Tozzi} et~al.}{2000}]{Tozzi2000}
{Tozzi} P.,  {Madau} P.,  {Meiksin} A.,   {Rees} M.~J.,  2000, \mn@doi [\apj]
  {10.1086/308196}, \href {http://adsabs.harvard.edu/abs/2000ApJ...528..597T}
  {528, 597}

\bibitem[\protect\citeauthoryear{{Verhamme}, {Orlitov{\'a}}, {Schaerer}  \&
  {Hayes}}{{Verhamme} et~al.}{2015}]{Verhamme2015}
{Verhamme} A.,  {Orlitov{\'a}} I.,  {Schaerer} D.,   {Hayes} M.,  2015, \mn@doi
  [\aap] {10.1051/0004-6361/201423978}, \href
  {http://adsabs.harvard.edu/abs/2015A%26A...578A...7V} {578, A7}

\bibitem[\protect\citeauthoryear{{Voytek}, {Natarajan}, {J{\'a}uregui
  Garc{\'{\i}}a}, {Peterson}  \& {L{\'o}pez-Cruz}}{{Voytek}
  et~al.}{2014}]{Voytek2014}
{Voytek} T.~C.,  {Natarajan} A.,  {J{\'a}uregui Garc{\'{\i}}a} J.~M.,
  {Peterson} J.~B.,   {L{\'o}pez-Cruz} O.,  2014, \mn@doi [\apjl]
  {10.1088/2041-8205/782/1/L9}, \href
  {http://adsabs.harvard.edu/abs/2014ApJ...782L...9V} {782, L9}

\bibitem[\protect\citeauthoryear{{Worseck} et~al.,}{{Worseck}
  et~al.}{2014}]{Worseck2014}
{Worseck} G.,  et~al., 2014, \mn@doi [\mnras] {10.1093/mnras/stu1827}, \href
  {http://adsabs.harvard.edu/abs/2014MNRAS.445.1745W} {445, 1745}

\bibitem[\protect\citeauthoryear{{Wouthuysen}}{{Wouthuysen}}{1952}]{Wouthuysen1952}
{Wouthuysen} S.~A.,  1952, \mn@doi [\aj] {10.1086/106661}, \href
  {http://adsabs.harvard.edu/abs/1952AJ.....57R..31W} {57, 31}

\bibitem[\protect\citeauthoryear{{Xu}, {Wise}, {Norman}, {Ahn}  \&
  {O'Shea}}{{Xu} et~al.}{2016}]{Xu2016}
{Xu} H.,  {Wise} J.~H.,  {Norman} M.~L.,  {Ahn} K.,   {O'Shea} B.~W.,  2016,
  \mn@doi [\apj] {10.3847/1538-4357/833/1/84}, \href
  {http://adsabs.harvard.edu/abs/2016ApJ...833...84X} {833, 84}

\bibitem[\protect\citeauthoryear{{Yajima}, {Choi}  \& {Nagamine}}{{Yajima}
  et~al.}{2011}]{Yajima2011}
{Yajima} H.,  {Choi} J.-H.,   {Nagamine} K.,  2011, \mn@doi [\mnras]
  {10.1111/j.1365-2966.2010.17920.x}, \href
  {http://adsabs.harvard.edu/abs/2011MNRAS.412..411Y} {412, 411}

\bibitem[\protect\citeauthoryear{Yue, Ferrara  \& Xu}{Yue
  et~al.}{2016}]{Yue2016}
Yue B.,  Ferrara A.,   Xu Y.,  2016, \mn@doi [Monthly Notices of the Royal
  Astronomical Society] {10.1093/mnras/stw2145}, 463, 1968

\bibitem[\protect\citeauthoryear{{Zahn}, {Mesinger}, {McQuinn}, {Trac}, {Cen}
  \& {Hernquist}}{{Zahn} et~al.}{2011}]{Zahn2011}
{Zahn} O.,  {Mesinger} A.,  {McQuinn} M.,  {Trac} H.,  {Cen} R.,   {Hernquist}
  L.~E.,  2011, \mn@doi [\mnras] {10.1111/j.1365-2966.2011.18439.x}, \href
  {http://adsabs.harvard.edu/abs/2011MNRAS.414..727Z} {414, 727}

\bibitem[\protect\citeauthoryear{{van Haarlem} et~al.,}{{van Haarlem}
  et~al.}{2013}]{van_Haarlem2013}
{van Haarlem} M.~P.,  et~al., 2013, \mn@doi [\aap]
  {10.1051/0004-6361/201220873}, \href
  {http://adsabs.harvard.edu/abs/2013A%26A...556A...2V} {556, A2}

\makeatother
\end{thebibliography}




\appendix

\section{Flexibility of the functional form for LFs}\label{app:LFs}

\begin{figure*}
\begin{center}
\includegraphics[width=18cm]{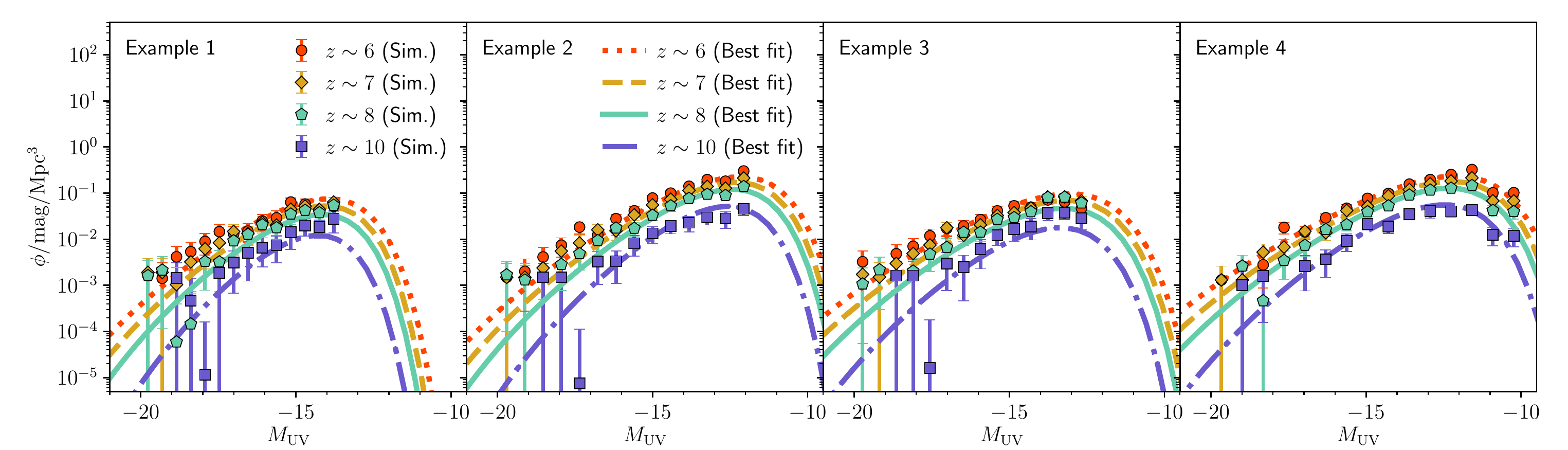}
\end{center}
\vspace{-3mm}
\caption{
LFs from hydrodynamic simulations are shown with points, together with the our best-fit model.  The hydrodynamic simulations were chosen to have good agreement with observed LFs at the bright end, but different trends at the faint end / high$-z$.  We note that our analytic model, based on the HMF, does not have any free parameters which regulate redshift evolution.
}
\label{fig:EMMA}
\end{figure*}

Here we quantify further the claim that our analytical model, based on the HMF, is flexible enough to fit ``reasonable" luminosity functions.  To do so, we make use of several hydrodynamic cosmological simulations, which form part of the on-going PRACE tier-0 project GAFFER.  The simulations were generated using the cosmological code, EMMA \citep{Aubert2015}, which includes a classical star formation recipe and supernova feedback [Deparis et al. (in prep)].  
As part of GAFFER, we perform many simulations varying parameters such as star formation efficiency, star formation density threshold and supernova efficiency. The simulations have a box length of $10\,{\rm Mpc}$ and resolve halo masses down to $10^8\,{\rm M_{\sun}}$.  They will be presented in an upcoming work,  Gillet et al. (in prep).

For our purposes here we take four simulation results which have among the best agreement with existing LF observations, but are different at the faint end and high redshifts at which we have little or no data.  We can thus test the ability of our analytical framework to capture diverse, yet physically reasonable LFs.

We run an MCMC of our model parameters using the LFs from EMMA as a mock observation and find maximum likelihood parameters. We include Poisson errors for the numbers of both dark matter halos and star particles, adding them in quadrature. Fig.~\ref{fig:EMMA} shows LFs generated from the simulation and the corresponding LFs with our maximum likelihood parameters. We find that in all four examples, our model is sufficiently flexible to fit the simulated LFs reasonably well.

\section{21-cm power spectra}\label{app:Mock}

The light-cone of the mock 21-cm observation is generated from $500\,{\rm Mpc}$ side length co-eval cubes with a $256^3$ grid, smoothed down from a high-resolution density field of $1024^3$.  To compute the mock 21-cm PS we follow the same approach as \cite{Greig2018}. We split the light-cone into equal comoving distance boxes and calculate the 21-cm PS (equation~\ref{eq:PS}) for each separate box. For the MCMC samples we generate a light-cone from $250 {\rm Mpc}$ side length co-eval cubes with a $128^3$ grid, smoothed down from a $512^3$ density field, but using different initial conditions. The mock observation is split into equal comoving distance boxes equivalent to the box length of the sampled boxes (i.e. $250\,{\rm Mpc}$). Then, we compute the 21-cm PS from the same comoving scale for both the mock observation and the MCMC samples. Since the light-cones extend from $z=6$ ($\sim200\,{\rm MHz}$) to $z=26.8$ ($\sim50\,{\rm MHz}$), this generates a total of 12 independent 'chunks'. 

\begin{figure*}
\begin{center}
\includegraphics[width=17cm]{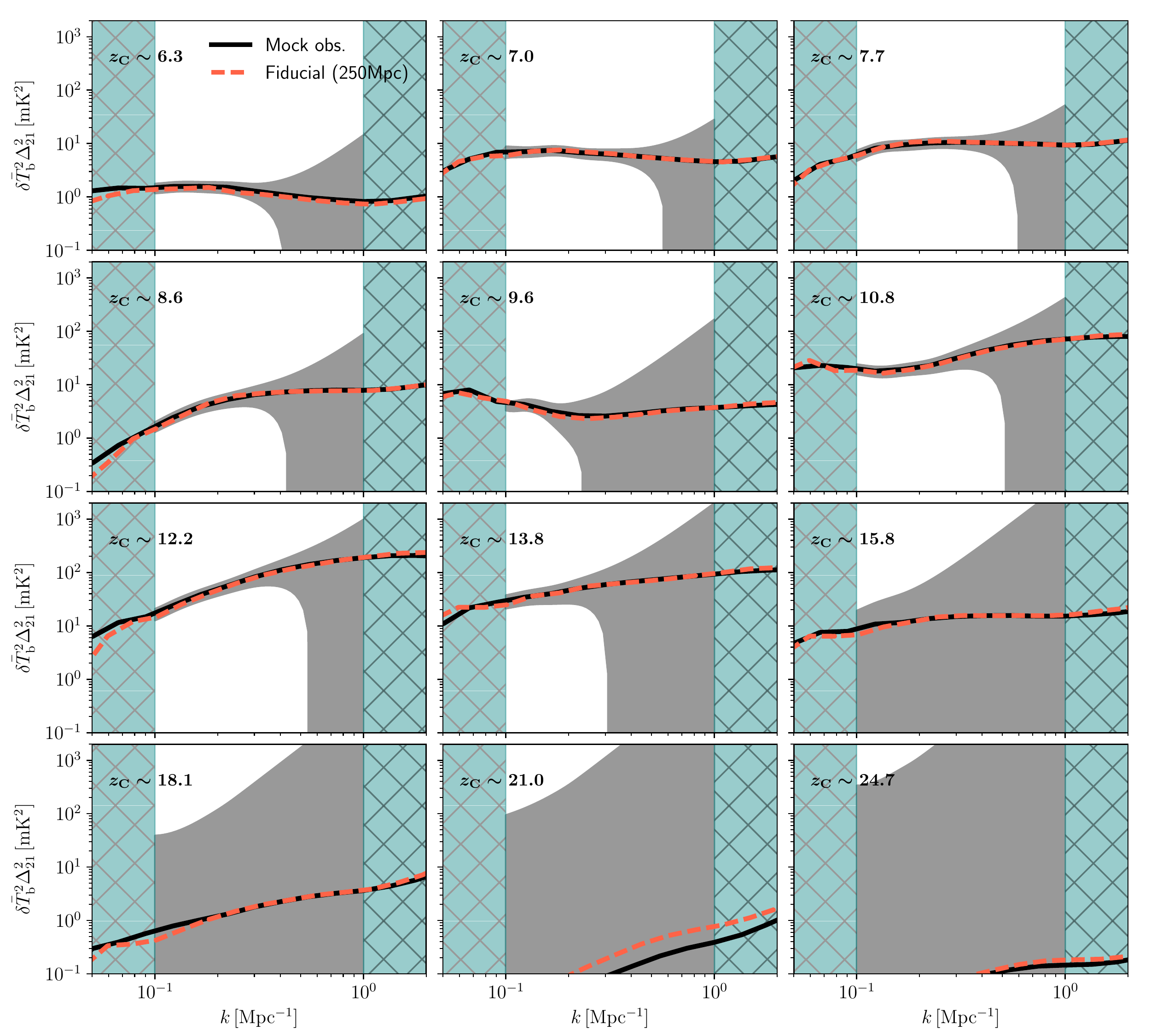}
\end{center}
\vspace{-3mm}
\caption{
The 21-cm PS from the mock observation (solid lines), and corresponding $1  \sigma$ errors assuming a 1000h observation with HERA331.  Dashed lines represent the MCMC sample with the fiducial parameters, but from a different random seed.  Hatched regions represent $k$ modes outside of our fitting range of $k=0.1-1\,{\rm Mpc^{-1}}$. $z_{\rm C}$ denotes the central redshift of each {\it 'chunk'} of the light-cone.
}
\label{fig:MockPS}
\end{figure*}

Fig.~\ref{fig:MockPS} shows the 21-cm PS for the mock observation generated (solid lines) and a sample 21-cm PS generated from the $250\,{\rm Mpc}$ box, using the same fiducial parameters but using a different initial seed. The shaded regions show the estimated noise corresponding to the mock 21-cm PS. We assume HERA for the noise estimate with a core design consisting of 331 dishes and a 1000 h observation. In each panel we denote the central redshift for each {\it 'chunk'} of the light-cone.

\section{Ionizing emissivity}\label{app:Emissivity}

Another potentially important data set on the high-$z$ source population is the ionizing emissivity as estimated from the Lyman $\alpha$ forest.  Here we study how this additional data set can further inform our models (c.f. \citealp{Choudhury&Ferrara2006,Kuhlen2012,Mitra2013, Bouwens2015}).

The ionizing emissivity is estimated by using the opacity measured from high-$z$ quasar spectra.  Post reionization, the optical depth in the IGM scales roughly as $\tau_{\rm Ly_{\alpha}} \propto T^{-0.7}\Delta^2_{\rm b}/\Gamma$, where $T$ is the gas temperature, $\Delta_{\rm b}$ is the gas density in units of the cosmic mean and $\Gamma$ is the photoionization rate.  The ionizing emissivity, $\epsilon$, can then be estimated using the post-reionization relation $\Gamma \propto \lambda_{\rm mfp}^{912} \epsilon$, where $\lambda_{\rm mfp}^{912}$ is the mean free path of ionizing photons.   This emissivity can then be directly compared to our model prediction from equation \ref{eq:nion}.

This procedure is non-trivial for several reasons.  Firstly, the Lyman $\alpha$ forest  is only sensitive enough at $z \lsim $ 5 to provide a reasonable estimate of the emissivity. The galaxies at these post-EoR redshifts could evolve beyond what is expected during the first billion years, due to feedback processes.  Thus they are not the same population that we are modeling.  More importantly, although galaxies are expected to dominate the EoR, it is likely that the contribution of AGN ramps up soon afterwards and thus cannot be ignored at these lower redshifts \citep[e.g.][]{Haardt&Madau2012,Chardin2015,Mitra2018}. We therefore take the emissivity estimates at $z\sim 5$ as {\it upper limits} to our galaxy emissivities.

Additionally, as explained above, we require knowledge of the IGM temperature, density and mean free path in order to estimate the emissivity from the forest.  This can be tricky by $z\sim 5$, with the mean free path being especially difficult to constrain to high precision.  Moreover, spatial fluctuations in these quantities can bias estimates.

Here we explore the utility of IGM emissivity upper limits for our parameter study, using the estimates from \cite{D'Aloisio2018}.  These authors estimated the ionizing emissivity at $4.8<z<5.8$ based on the measurement of $\tau_{\rm Ly_{\alpha}}$ by \cite{Becker2015}. They post-processed simulations to compute a spatially varying photoionization rate, $\Gamma$, and rescaled it to fit the observed $\tau_{\rm Ly_{\alpha}}$ under the assumption $\lambda_{\rm mfp}^{912}({\mathbf x})\propto \Gamma^{2/3}({\mathbf x})/\Delta({\mathbf x})$, where $\Delta({\mathbf x})$ is the local matter density and $\lambda_{\rm mfp}^{912}$ is the mean free path of ionizing photons. This rescaling provides the ionizing emissivity, $\epsilon_{912}$, with the relation $\Gamma \propto \lambda_{\rm mfp}^{912} \epsilon_{912}$. They use three models for the mean free path, which they refer to as fiducial, intermediate and short. The fiducial one is consistent with the mean free path measurements of \cite{Worseck2014} which are at $z \leq 5.2$, though \citet{D'Aloisio2018} argue this might be an overestimate due to a bias from including the proximity zone in the mean free path calculation.

\begin{figure}
\begin{center}
\includegraphics[width=9cm]{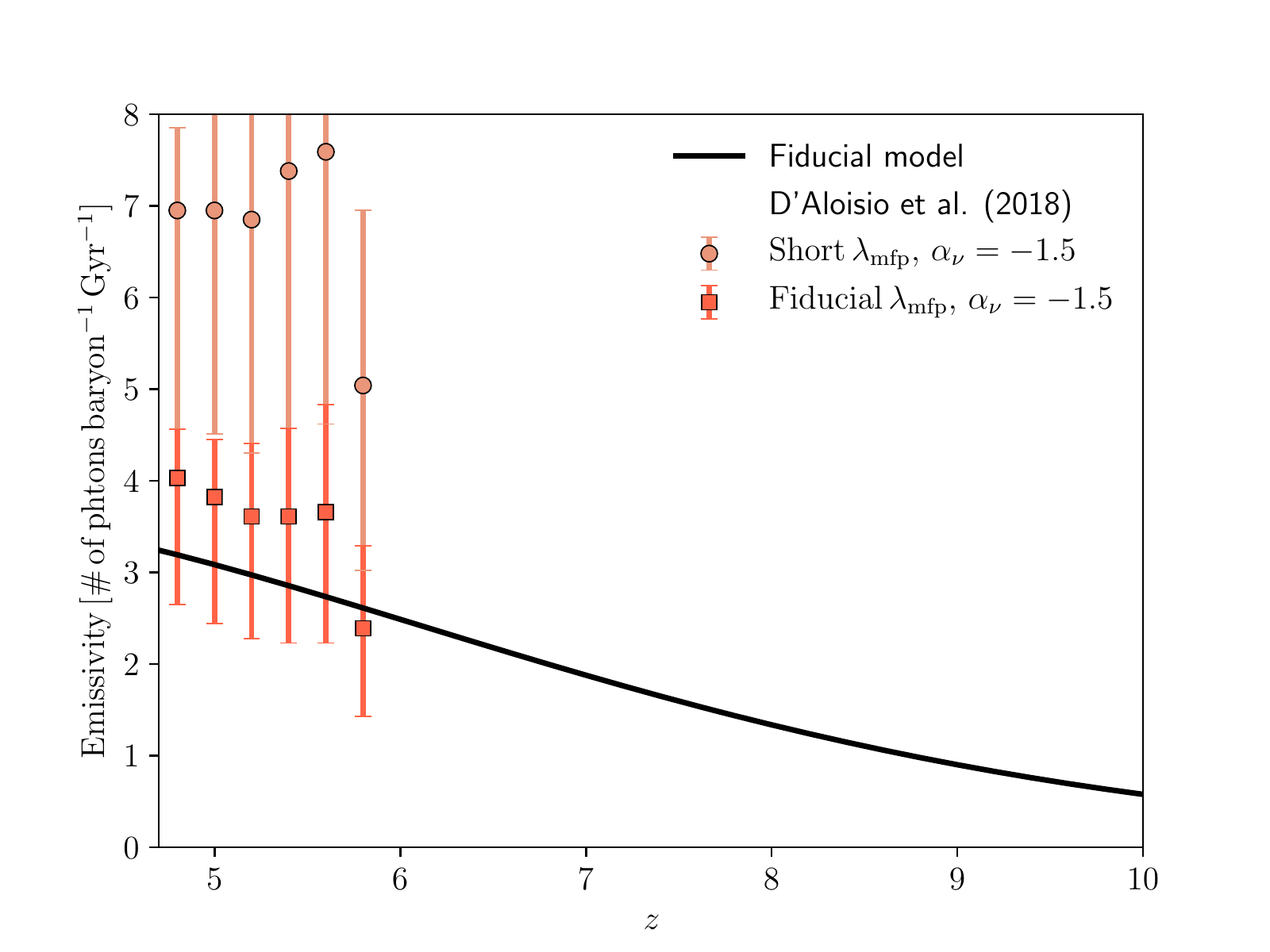}
\end{center}
\vspace{-3mm}
\caption{
Redshift evolution of the ionizing emissivity. Solid line represents the prediction of our fiducial model. Squares and circles with error bars represent the measured emissivity by \protect\cite{D'Aloisio2018} with their fiducial mean free path and short mean free path, respectively. We note that to convert units we assume the ionizing specific luminosity follows a power-law, $L_{\rm \nu}\propto \nu^{\alpha_{\rm \nu}}$, and adopt $\alpha = -1.5$ which is similar quantity in \protect\cite{Lusso2015}
}
\label{fig:Emissivity}
\end{figure}

The resulting estimates of the ionizing background in the fiducial and short mean free path models are shown as points with error bars in Figure~\ref{fig:Emissivity}.  To convert the emissivity to number of photons per baryon per Gyr, we assume the  specific emissivity provided by \citet{D'Aloisio2018} follows a power-law, $L_{\rm \nu}\propto \nu^{\alpha_{\rm \nu}}$, and adopt $\alpha = -1.5$, consistent with \cite{Lusso2015} (see also e.g. \citealp{McQuinn2011,D'Aloisio2018b}). With the solid curve, we also show the emissivity from our fiducial parameter set, used to generate the mock 21-cm signal.  This emissivity is roughly consistent with the fiducial mean free path model of \citet{D'Aloisio2018}.

Given the uncertainties in these estimates, how constraining is the emissivity for our parameter space?  To quantify this, we use the fiducial mean free path estimates  of \cite{D'Aloisio2018} (which are lower and thus more constraining) at $z\sim 5.4$ and $5.8$ as upper limits (allowing for an additional AGN contribution as discussed above).  Specifically, we take a flat prior for values lower than the points, and then a one-sided Gaussian decreasing for higher emissivities with the sigma reported by these authors.  We then re-run our MCMC for the UV astrophysical parameters, with and without this additional data set.

\begin{figure*}
\begin{center}
\includegraphics[width=18cm]{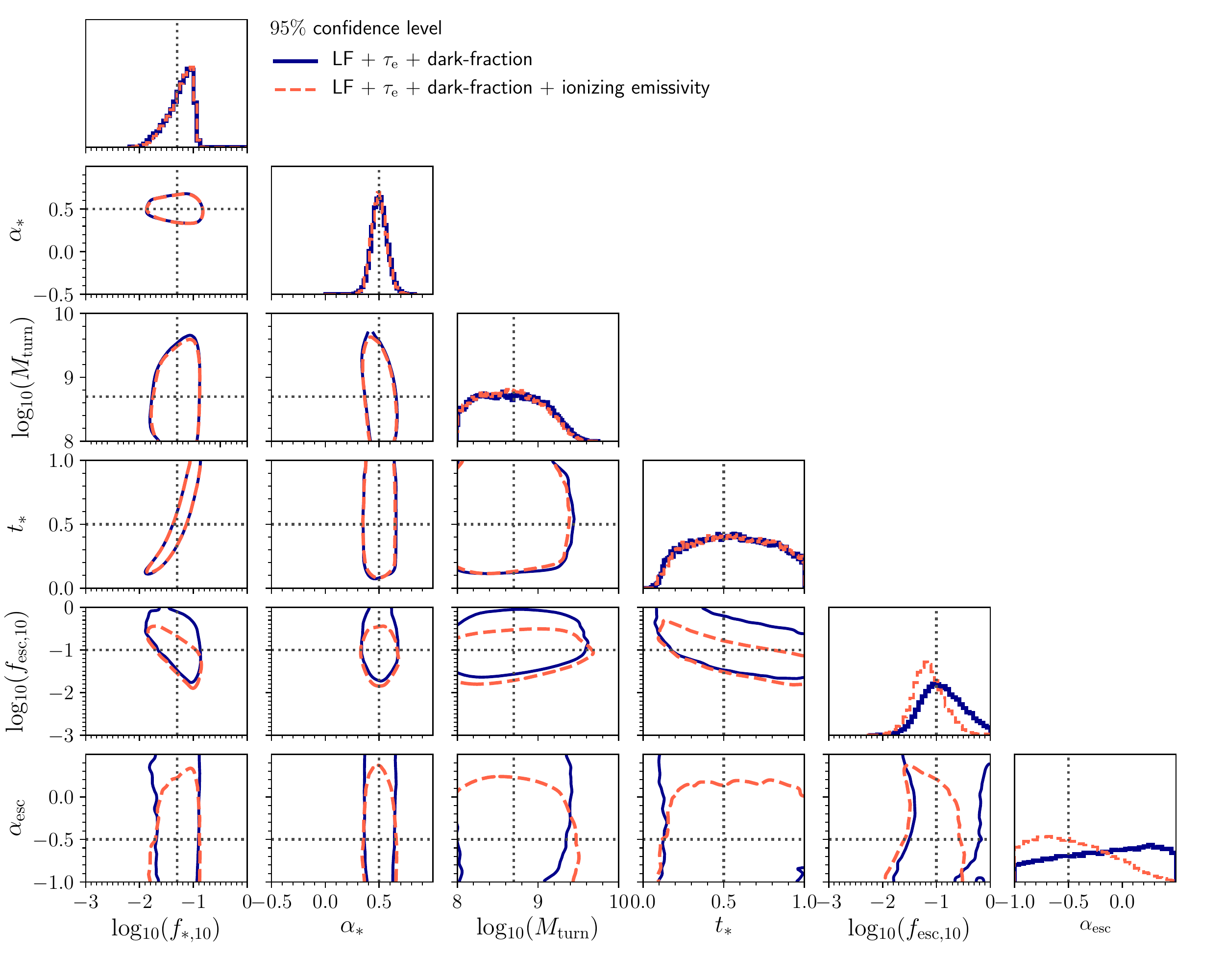}
\end{center}
\vspace{-3mm}
\caption{
Marginalized joint posterior distributions for UV galaxy properties with and without a prior on the emissivity. Solid (blue) and dashed (red) lines represent $95$ per cent confidence levels for constraints using LF + $\tau_{\rm e}$ + the dark fraction (same as in Fig. \ref{fig:contour}), and when additionally using the ionizing emissivity. The minor relative differences between these curves demonstrate that the ionizing emissivity currently has little additional constraining power for our model.
}
\label{fig:Triangle_emissivity}
\end{figure*}

The resulting constraints are shown in Figure~\ref{fig:Triangle_emissivity}.
Even for the conservatively strong prior of using the fiducial emissivity estimates (as opposed to the higher ones provided by \citealt{D'Aloisio2018}), the constraints are quite comparable to those already presented in Fig. \ref{fig:contour}. However, we see that a $1\sigma$ percentage error for the escape fraction is reduced from $\sim 40$ per cent to $\sim 25$ per cent, while the power-law of the escape fraction is still not constrained.


\bsp	
\label{lastpage}
\end{document}